\documentclass{aa}

\usepackage{graphicx}
\usepackage{natbib}
\usepackage{comment}
\bibpunct{(}{)}{;}{a}{}{,}

\newcommand{\mic}{\mathrm{\mu s}}

\begin{document}

\title{Pulsar microstructure and its quasi-periodicities with
the S2 VLBI system at a resolution of 62.5 nanoseconds}

\titlerunning{Pulsar microstructure and its quasi-periodicities}
\authorrunning{M.V. Popov et al.}

\author{M.V. Popov\inst{1}
	\and N. Bartel\inst{2}
	\and W.H. Cannon\inst{2,3}
        \and A.Yu. Novikov\inst{3}
        \and V.I. Kondratiev\inst{1}
	\and V.I. Altunin\inst{4}
       }

\offprints{M.V. Popov, \\
\email{mpopov@asc.rssi.ru}}

\institute{Astro Space Center of the Lebedev Physical Institute,
           Profsoyuznaya 84/32, Moscow, 117997 Russia
           \and
	   York University, Department of Physics and Astronomy,
	   4700 Keele Street, Toronto, Ontario M3J 1P3 Canada
           \and
	   Space Geodynamics Laboratory/CRESTech,
	   4850 Keele Street, Toronto, Ontario M3J 3K1 Canada
	   \and
	   Jet Propulsion Laboratory, 4800 Oak Grove Drive,
           Pasadena, CA 91109 U.S.A.
           }

\date{Received ~~~~~~~~~~~~~~~ / Accepted ~~~~~~~~~~~~~~~}

\abstract{
We report a study of microstructure and its quasi-periodicities of three
pulsars at 1.65 GHz with the S2 VLBI system at a resolution of 62.5~ns,
by far the highest for any such statistical study yet.
For PSR B1929+10 we found in the average cross-correlation function (CCF)  
broad microstructure with a characteristic timescale of $95\pm 10~\mic$
and confirmed microstructure with characteristic timescales between 100
and $450~\mic$ for PSRs B0950+08 and B1133+16.
On a finer scale PSRs {B0950+08}, \object{B1133+16} (component II)
and \object{B1929+10} show narrow
microstructure with a characteristic timescale in the CCFs of $\sim 10~\mic$,
the shortest found in the average CCF or
autocorrelation function (ACF) for any pulsar, apart perhaps for the Crab pulsar.
Histograms of microstructure widths are skewed heavily toward shorter timescales
but display a sharp cutoff. The shortest micropulses have widths between 2 and $7~\mic$.
There is some indication that the timescales of the broad, narrow,
and shortest micropulses are, at least partly, related to the widths of the
components of the integrated profiles and the subpulse widths. If the shortest micropulses
observed are indeed due to beaming then the ratio, $\gamma$, of the relativistic
energy of the emitting particles to the rest energy is about 20\,000, independent
of the pulsar period. We predict an observable lower limit for the width of micropulses
from these pulsars at 1.65~GHz of $0.5~\mic$. If the short micropulses are instead
interpreted as a radial modulation of the radiation pattern, then the associated
emitting sources have dimensions of about 3 km in the observer's frame. For PSRs
\object{B0950+08} and \object{B1133+16} (both components) the
micropulses had a residual dispersion delay over a 16~MHz frequency difference
of $\sim 2~\mic$ when compared to that of average pulse profiles over
a much larger relative and absolute frequency range. This residual delay is
likely the result of propagation effects in the pulsar magnetosphere that contribute to
limiting the width of micropulses. No nanopulses or unresolved pulse spikes were
detected. Cross-power spectra of single pulses show a large range of complexity with
single spectral features representing classic quasi-periodicities  and broad and overlapping
features with essentially no periodicities at all. Significant differences were found
for the two components of \object{PSR~B1133+16} in every
aspect of our statistical analysis of micropulses and their quasi-periodicities.
Asymmetries in the magnetosphere and the hollow cone of emission above the polar cap of the
neutron star may be responsible for these differences.
}

\maketitle

\section{Introduction}

Pulsar radio emission originates in a region of extremely small size, most
likely from charged particles in the magnetosphere traveling along the diverging dipole
magnetic field lines above the polar cap of a neutron star~\citep[e.g.,][]{ruderman1975,
arons1983}. This emission can be seen by an external observer only during short successive
time windows separated by the neutron star's rotation period. Within such a window, the radio
signals recorded at different times are therefore related to different reference points
in the pulsar magnetosphere which are either separated longitudinally across the beam or
radially along the beam or by a combination of both. Correspondingly, the observed intensity
fluctuations can be either caused by a longitudinal modulation of the radiation pattern over
the cross section of the polar magnetic field lines or by a radial modulation of the
radiation pattern along the opening polar magnetic field lines, or, again, by a combination of
both. The longitudinal modulation is most likely related to the stationary geometry of the
emission beam fixed to each of the poles of the rotating neutron star. The radial modulation
is likely related to plasma bunching and linked to the elementary emission mechanism. In
this model the spectrum of the radio emission is a function of the radial distance from the
neutron star, and the beam width is frequency dependent. High frequency radiation is emitted
closer to the neutron star and the beam is narrower, low frequency radiation is emitted
further out and the beam is broader, reflecting the opening of the polar magnetic field lines. The
study of pulsar intensity fluctuations has largely the goal of probing on the one hand the
geometrical characteristics of the pulsar emission beam and its underlying magnetospheric
structure and on the other hand the elementary emission mechanism.

Pulsar radio emission is known to exhibit fluctuations over a broad range of
timescales. Average pulse profiles can have up to seven components~\citep{kramer1994} and
together with their frequency dependent widths reflect best the underlying geometrical
structure of the magnetosphere~\citep{rankin1983}. Every individual pulse is composed of one or
several separate subpulses. In general, the subpulses fluctuate strongly within a
single pulse and from pulse to pulse but have stationary characteristics and characteristic
widths computed from their autocorrelation functions that are $96\%$
correlated with the width of the strongest component of the average pulse
profile~\citep{bartel1980}. Like the average pulse profiles they most likely
also reflect the geometrical structure of the magnetosphere~\citep{bartel1980}.
The subpulses in their turn are often composed of micropulses or microstructure
with typical timescales of about hundred to
a few hundred microseconds, or several tenths of a degree in pulsar
longitude~\citep[e.g.,][]{hankins1972, kardashev1978}.
In a few cases still much faster but well resolved individual fluctuations were
recorded, for instance with a timescale down to $2.5~\mic$ for
\object{PSR~B1133+16}~(\citealt{bartel1982}, see also~\citealt{bartel1978}), the fastest
fluctuations found for any pulsar apart from the Crab pulsar. For the latter
pulsar sporadic giant pulses were observed which were still unresolved
at a time resolution of 10~ns~\citep{hankins2000}.

For the broader micropulses, quasi-periodic structures were found in the very
first studies of single pulses with sufficiently high time resolution. \citet{hankins1971}
found many examples of regularly spaced micropulses with periods
of 300 to $700~\mic$ for \object{PSR~B0950+08} at a frequency of
111.5~MHz. \citet{backer1973}, \citet{boriakoff1976} and \citet{cordes1976a}
have shown that \object{PSR~B2016+28} has quasi-periodic microstructure with periods
ranging from 0.6 to 1.1~ms at a frequency of 430~MHz. \citet{soglasnov1981, soglasnov1983}
analyzed the  statistics of quasi-periodicities for PSRs \object{B0809+74} and
\object{B1133+16} at 102.5~MHz. \citet{cordes1990} studied five pulsars
(including PSRs \object{B0950+08} and \object{B1133+16}) with quasi-periodic
microstructure at several radio frequencies. They concluded that there are no
preferred periods for quasi-periodicities that are intrinsic to a given pulsar
and that there is  no frequency dependence of the micropulse
width and the  characteristic period of the quasi-periodicity.

Lange et al. (1998) studied seven bright pulsars,
including our three, at 1.41 and 4.85~GHz with a time resolution between
$7~\mic$ and $160~\mic$. They did not find notable differences
of microstructure parameters at different frequencies.

It is the topic of this paper to investigate micropulses and their
quasi-periodicities and to help to understand whether they
reflect the longitudinally modulated emission pattern and the geometry of the
magnetosphere~\citep[e.g.,][]{benford1977} or instead are more effected by the
radially or intrinsically temporally modulated pulsar emission
pattern~\citep[e.g.,][]{hankins1972, cordes1981}. The  shortest micropulses observable in a pulsar
in particular may harbor essential clues about the nature of the emission
fluctuations in pulsars.

To achieve a high time resolution one must digitally record the pulsar signal
before detection with subsequent dispersion removal processing as
originally described by~\citet{hankins1971}.
Previous studies were based mainly on observations made with
a time resolution of several microseconds to several tens of microseconds.
A review of microstructure research is given by~\citet{hankins1996}.
In this paper we present a statistical analysis of the properties of
microstructure for PSRs \object{B0950+08},
\object{B1133+16}, and \object{B1929+10} at 1\,650~MHz
with a time resolution of 62.5~ns, the most extensive such analysis yet for any
pulsar and with one of the  highest time resolutions ever used.

\begin{table*}

\caption{Pulsar characteristics and fixed  parameters: P is the pulsar period;
DM is the  dispersion measure for a dispersion constant:
$\alpha_\mathrm{d} = 2.41\times 10^{-16}~\mathrm{cm^{-3}pc\cdot s}$;~
$\delta t_\mathrm{cal}$ is the computed time delay between
the two frequency channels based on DM, the standard errors are $<0.1~\mic$;
$N$ is the approximate total number of pulsar periods observed;
$N_\mathrm{p}$ is the number of selected strong pulses used to compute the CCFs.
For \object{PSR~B1929+10} a distinction is made between the  number used to
compute the average CCF and the number (in parentheses) used to compute the
individual CCFs for the analysis of individual pulses;
$N_\mathrm{\mu}$ is  the number of microstructure features revealed;
$N_\mathrm{\mu-QP}$ is the number of detected quasi-periodicities;
$N_\mathrm{no-\mu}$ is the number of
smooth structureless pulses; $T$ is the number of samples for the duration
of the pulse window used to calculate the CCF;
{\em SNR} is the signal-to-noise ratio of the mean
pulse intensity in the ON-pulse window averaged over all  pulses selected for processing.
It is a measure of the relative increase of the antenna temperature in the ON-pulse window.
For the specific definition of {\em SNR}, see text.}
\label{psrpar1}

\begin{flushleft}
\begin{tabular}{lcccccccccc}
\hline
{\Large\strut}PSR B                  & P                                      & $\mathrm{DM}$                          &
$\delta t_\mathrm{cal}$              & $N$                                    & $N_\mathrm{p}$                         &
$N_\mathrm{\mu}$                     & $N_\mathrm{\mu-QP}$                    & $N_\mathrm{no-\mu}$                    &
$T$                                  & {\em SNR}   \\
                                     & (s)                                    & $\mathrm{(pc/cm^3)}$                   &
$(\mic)$                             &                                        &                                        &
                                     &                                        &                                        &
                                     &         \\
\hline
{\Large\strut}0950+08                & 0.253                                  & $2.9702^\mathrm{a}$                    &
$\phantom{1}89.1$                    & 14\,000                                & $\phantom{2}225\phantom{~(492)}$       &
$\phantom{1\,}716$                   & 568                                    & $\phantom{5}1$                         &
131\,072                             & 0.29    \\
1133+16 (I)                          & 1.188                                  & $4.8471^\mathrm{a}$                    &
145.4                                & $\phantom{1}3\,000$                    & $\phantom{2}240\phantom{~(492)}$       &
$\phantom{1\,}362$                   & 544                                    & 41                                     &
524\,288                             & 0.63    \\
1133+16 (II)                         &                                        &                                        &
145.4                                &                                        & $\phantom{2}132\phantom{~(492)}$       &
$\phantom{1\,}165$                   & 134                                    & 51                                     &
524\,288                             & 0.33    \\
{\large\strut}1929+10                & 0.226                                  & $3.1760^\mathrm{b}$                    &
$\phantom{1}95.3$                    & 15\,500                                & $\phantom{2}998~(492)$                 &
1\,093                               & 602                                    & 14                                     &
262\,144                             & 0.28    \\
\hline
\multicolumn{3}{l}{{\large\strut}${}^\mathrm{a}$~\citet{phillips1992}} &
&                      &                       &
&                      &                       &           &       \\
\multicolumn{3}{l}{${}^\mathrm{b}$~\citet{manchester1972}}             &
&                      &                       &
&                      &                       &           &       \\

\end{tabular}
\end{flushleft}

\end{table*}

\section{Observations}

The observations were made with the NASA Deep Space Network 70-m DSS43 radio
telescope at Tidbinbilla, Australia. PSRs \object{B0950+08} and \object{B1133+16} were
observed on 10~May~2000  and \object{PSR~B1929+10} on 24~April~1998.

The data were recorded continuously with the S2 VLBI system~\citep{cannon1997,
wietfeldt1997} in the 2-bit sampling mode in the lower sideband from 1\,634 to 1\,650~MHz
and the upper sideband from 1\,650 to 1\,666~MHz. Left circular polarization
was recorded for both frequency channels. The observations were made \emph{in
absentia} which is more typical for VLBI observations. In general, pulsar observations with the
S2 VLBI system can be made at any of the  $\sim 30$ radio telescopes worldwide
which are equipped with such a system, in the same way VLBI observations are made without
the need for the investigator's presence. In effect, a dedicated pulsar backend at
the observing station is replaced with a software package on the workstation at the
investigator's home institution.

\section{Data reduction}

The tapes were shipped to Toronto and played back through the
S2 Tape-to-Computer Interface (S2-TCI) at the Space
Geodynamics Laboratory (SGL) of CRESTech on the campus of York University.
The S2-TCI system transfers the baseband-sampled pulsar data to computer files
stored on hard disks and makes the data available for off-line analysis by a
SUN Workstation computer. A similar use of the S2-TCI system for processing
observations of \object{PSR~J0437$-$4715} was reported by~\citet{kempner1997}.

The S2-TCI system enabled us to transfer the data stream to disk in a
piecewise manner, the size or duration of each piece of data  stream being
limited by the disk storage capacity. In general, we transferred a set of about 10 minutes
of data to disk at a time. In a first step we detected the recorded signal
by squaring and averaging it  with a time constant of about $100~\mic$. Then we
determined the  phase of the ON-pulse window  and selected strong pulses. 
For the selection we computed the signal-to-noise ratio, 
$\mathit{SNR}=\frac{\langle I_\mathrm{on}\rangle}{\langle I_\mathrm{off}\rangle}$,
where $\langle I_\mathrm{on}\rangle$ is the mean
intensity in the ON-pulse window after subtraction of the mean intensity in the OFF-pulse window, 
$\langle I_\mathrm{off}\rangle$. The {\em SNR} so defined corresponds  to the relative
increase of the  antenna temperature in the ON-pulse window and is therefore relatively small 
even for strong pulses. We used $\mathit{SNR}=0.1$
as a threshold for pulse selection.
The approximate total
number of pulses, $N$, observed and the number of selected pulses,
$N_\mathrm{p}$, are listed for each pulsar in Table~\ref{psrpar1}. This approach
reduced the amount of data by a factor of several hundred and enabled us to
carry out the subsequent signal processing more efficiently.

Having determined the pulse windows and selected strong pulses we further
processed the recorded (raw) signal by decoding the signal amplitude sampled in two bits.
Two-bit sampling of the amplitude of Gaussian random noise is widely used in
VLBI observations.
The decoding is generally done using four levels with
integer values equal to $-3$, $-1$, $+1$, $+3$~\citep{thompson1988}.
These values reasonably represent the signal while the threshold level
where the sampler switches from 1 to 3 and from $-1$ to $-3$ is equal to
the current root-mean-square (rms) deviation ($+1\sigma$ and $-1\sigma$, respectively)
of the signal. In order to preserve this condition during observations, the S2 VLBI data
acquisition system (S2-DAS) has an automatic gain control (AGC). For pulsar
observations with the S2-DAS it is preferable to switch off the AGC and instead
use the manual gain control with the gain fixed,
or if left switched on, to choose a sufficiently long time constant for the AGC
loop. Either option has the advantage of preventing the sampler from experiencing sudden gain
discontinuities inside a pulse window. For our observations the AGC was inadvertently left
switched on, but fortunately the gain was found to be constant inside
the selected pulse windows in the majority of cases. Therefore the threshold
level could relatively easily be adjusted during the analysis after the observations
to reflect the larger voltage variations in the ON-pulse window. We changed
the decoding values through the data records from $\pm1$ and $\pm3$ to real values
that correspond to the current $\pm1 \sigma$ levels
in accordance with the technique developed by~\citet{jenet1998}.
We computed the new levels from the quasi-instantaneous rms values of
subsequent portions of the data records, each  $100~\mic$ long,
to approximately match the dispersion smearing time across the 16-MHz bandwidth.

The next step in our data processing routine was the removal of the dispersion
caused by the interstellar medium.
The predetection dispersion removal technique itself~\citep{hankins1971}
consists of a Fourier transform of the decoded signal followed first by
amplitude corrections for the generally non-uniform
receiver frequency bandpass and phase corrections for the dispersion delay, and
then by an inverse Fourier transform back to the time domain. In particular, for the
phase corrections of the dispersion delay, $\delta\Phi(\nu)$, at the observing frequency, $\nu$,
we used

\begin{center}
\begin{equation}
\label{dedisp}
\delta\Phi(\nu)=\delta\Phi(\nu_0+\Delta\nu)=\frac{2\pi}{\alpha_\mathrm{d}}\cdot
\frac{\mathrm{DM}}{\nu_0^3}\Delta\nu^2\left(1-\frac{\Delta\nu}{\nu_0}\right),
\end{equation}
\end{center}
\noindent
where $\Delta\nu\ll\nu_0$, DM is the dispersion measure, and
$\alpha_\mathrm{d}$ is the dispersion constant (see caption of Table 1).
\noindent
Equation~(\ref{dedisp}) was derived from~\citet{hankins1975} by conserving
only quadratic and cubic terms in their equation (5). The next term would give a
correction of only about 10~ns for the dispersion smearing
over the 16-MHz bandwidth at 1650~MHz for a typical value of DM=3.0 pc/cm$^3$ for our
pulsars, which is only a sixth of our time resolution.
Also, we did not need to apply a Doppler shift correction
for DM due to the Earth's motion, since even for the largest possible velocity
component of $30\!~\mathrm{km/s}$ such a correction
would be limited to a smearing across the 16-MHz bandwidth of only about 40 ns
for any of our pulsars which is smaller than our time resolution.
The values of DM used for the three pulsars are listed in Table~\ref{psrpar1}.
After the amplitude and phase corrections we set to zero all
negative frequency components in the spectrum and then
computed the inverse Fourier transform.
The sum square of the
real and imaginary components of the complex time series represents
the dispersion-free detected signal of the pulsar~\citep{born_n_wolf}.

Finally, we calibrated the intensities. Their scale is based on the system 
temperature which was recorded in a log file. Noise fluctuations of the unsmoothed
detected signal of $1\sigma$ in the OFF-pulse window correspond to 50~Jy. Dynamically changing
 the conversion threshold levels during the ON-pulse and OFF-pulse 
windows permitted us to avoid a possible bias in the decoding of the two-bit samples.

\section{Results}
\label{res}
\subsection{Individual pulses} \label{ind_pulse}

\begin{figure}[hbt]

\includegraphics[width=9cm,height=9cm] {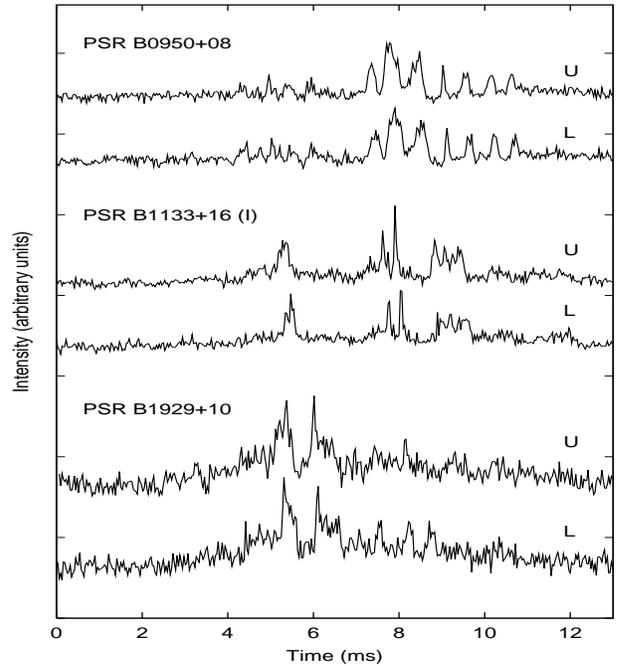}

\vskip 2mm

\caption{An example of a strong pulse at 1\,650~MHz (U) and 1\,634~MHz (L)
for each of the three pulsars.  The dispersion smearing was
removed, but the
dispersion delay
between frequency channels was not removed. The intensity fluctuations were
smoothed over a 32-$\mic$ time interval.}

\label{pulses}

\end{figure}

\noindent
In Figure~\ref{pulses} we display one example of a strong pulse for the
upper~(U) and lower~(L) sideband for each of the three pulsars.
The pulses are shown with a time resolution of $32~\mic$. It can  be seen that the
pulses vary in intensity on a timescale down to almost the resolution interval.
The intensity variations are strongly correlated between the upper and lower frequency
bands.

\begin{figure}[hbt]

\includegraphics[width=8cm,height=8cm] {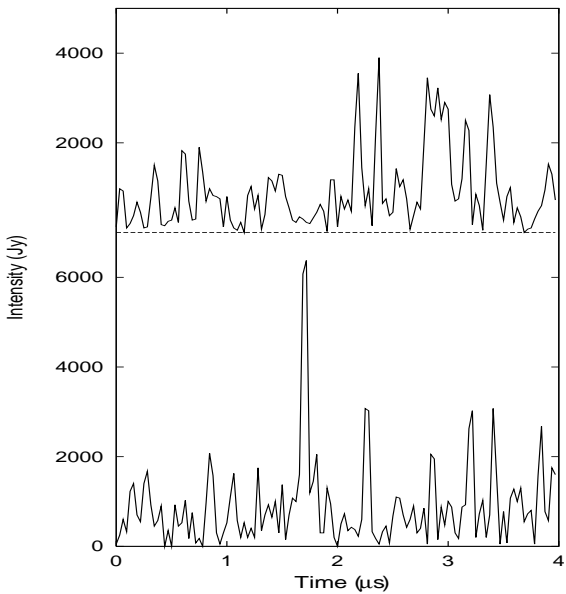}

\vskip 2mm

\caption{A short portion of a very strong pulse from  \object{PSR~B1133+16}
plotted with a 62.5~ns time  resolution. The upper curve corresponds
to the 1\,650~MHz frequency channel, and the lower one corresponds to
the 1\,634~MHz frequency channel. The dispersion smearing and
time delay between the frequency channels were removed. The original 2-bit
quantization of the voltages is not visible because of the dispersion removal.}

\label{giant}

\end{figure}

More rapid variations also appear to be present on first sight. For example,
one very strong pulse of \object{PSR~B1133+16} has a mean
flux density averaged over the pulse window of about 500~Jy, while the
strongest  isolated spike observed in the lower sideband
has a flux density in excess of 6\,000~Jy. A short 4-$\mic$~portion of this
pulse
is shown in
Figure~\ref{giant} with the original time resolution of 62.5~ns.
For such short intensity variations in time no correlation is apparent
between the data of the two sidebands
even if allowance is made for a possible variation of DM over a conceivable
range of values.
It is important to investigate whether these spikes and other strong unresolved
ones are  related to the physical emitters or are just statistically
insignificant noise fluctuations of much broader pulse structure that
increased the ON-pulse antenna temperature and led to the strong variations.

In the remainder of the paper we first discuss  the properties of the
microstructure with a width of $\sim 1$ to $\sim 500~\mic$. Then
we present a search for structure on sub-microsecond timescales,
and finally we report on a study of the statistics of microstructure
quasi-periodicities of the micropulses of the three pulsars.

\begin{table*}

\caption{Observed pulsar parameters:
$\delta t_\mathrm{obs}$ is the observed time delay between the two frequency
channels;
$\delta t_\mathrm{obs} - \delta t_\mathrm{cal}$ is the difference between the
observed
and computed time delay for the  two frequency channels;
$\tau_\mathrm{\mu-broad}$ and $\tau_\mathrm{\mu-narrow}$
are the characteristic timescales of the microstructure
determined from the average cross-correlation functions (see text);
$\tau_\mathrm{\mu-shortest}$ is the shortest width
found for the observed micropulses; $t_{1/2}$ is the FWHM of the average profile
or, in case of \object{PSR~B1133+16}~(I), of the component in the profile (taken
from~\citet{bartel1980}). For \object{PSR~B1133+16}~(II) no width is given in
~\citet{bartel1980}; the width is from \citet{kramer1994} interpolated between
his values of a FWHM of a Gaussian at 1.41 and 4.75 GHz to 1.65 GHz.
The parameter, $\tau_\mathrm{m}$, also from~\citet{bartel1980}, is a measure of
the typical subpulse width determined as the
$50\%$ width (ignoring the zero-lag spike) of the average ACF of single
pulses (for \object{PSR~B1133+16} $\tau_\mathrm{m} = 2.9 \pm 0.1$~ms, no
distinction is made between
the two components).}

\label{psrpar2}

\begin{flushleft}
\begin{tabular}{lccccccc}
\hline
{\Large\strut}PSR B                 & $\delta t_\mathrm{obs}$               & $\delta t_\mathrm{obs} - \delta t_\mathrm{cal}$                &
$\tau_\mathrm{\mu-broad}$           & $\tau_\mathrm{\mu-narrow}$            & $\tau_\mathrm{\mu-shortest}$                                   &
$t_{1/2}$                           & $\tau_\mathrm{m}$                 \\
                                    & $(\mic)$                              & $(\mic)$                                                       &
$(\mic)$                            & $(\mic)$                              & $(\mic)$                                                       &
$(\mathrm{ms})$                     & $(\mathrm{ms})$                   \\
\hline
{\Large\strut}0950+08               & $\phantom{1}87.2 \pm 0.3$             & $ -1.9 \pm 0.3 $                                               &
$135 \pm 5$                         & $14 \pm 3$                            & 7                                                              &
9.1                                 & 5.0                               \\
1133+16 (I)                         & $143.1 \pm 0.2$                       & $ -2.3 \pm 0.2 $                                               &
$430 \pm 30$                        & $\phantom{10}-\phantom{2}$            & 6                                                              &
$6.0 \pm 0.6 $                      & $\phantom{10}-\phantom{2}$        \\
1133+16 (II)                        & $143.1 \pm 0.2$                       & $ -2.3 \pm 0.2 $                                               &
$110 \pm 20$                        & $11 \pm 3$                            & 2                                                              &
$10.7 \pm 0.3$                      & $\phantom{10}-\phantom{2}$        \\
{\large\strut}1929+10               & $\phantom{1}95.2 \pm 0.1$             & $ -0.1 \pm 0.1 $                                               &
$\phantom{1}95 \pm 10$              & $\phantom{1}9 \pm 3$                  & 5                                                              &
$5.9\pm 0.2$                        & $3.4\pm 0.1 $                     \\
\hline

\end{tabular}
\end{flushleft}

\end{table*}

\subsection {Microstructure}
\subsubsection {The typical microstructure width from the average cross-correlation function}

\noindent
We determined the typical width of microstructure in a large number of single
pulses from the width of the central spike of the averaged cross-correlation
function (CCF). We chose CCFs for our analysis instead of the more widely
used autocorrelation functions (ACFs) to avoid the noise spike at zero lag of the latter
which could cause confusion in the detection of structure with the shortest timescales.
More precisely, for each pulsar we selected $N_\mathrm{p}$ (Table~\ref{psrpar1})
strong pulses, computed for each of them the CCF, $R_{1,2}(\tau)$,
between the signals, $I_1(t)$ and $I_2(t)$, in the lower and upper
sideband frequency channels, respectively, at lag $\tau$ with

\begin{center}
\begin{equation}
R_{1,2}(\tau)=[R_{1,1}(0)R_{2,2}(0)]^{-1/2}\sum_{t=1}^TI_1(t)I_2(t+\tau)~.
\label{eqsec}
\end{equation}
\end{center}
\noindent
Here $t$ is the sample number at a particular time within the single pulse,
$T$ (see Table~\ref{psrpar1}) is the total number of samples chosen for the
pulse window, and $I_1(t)$, $I_2(t)$ represent ON-pulse intensities with
the OFF-pulse mean levels subtracted.

We next averaged the CCFs for each pulsar and then corrected the averaged CCF
for the receiver noise by dividing it by the factor
$\epsilon$ (see Appendix~\ref{appA}):
\begin{equation}
   \epsilon = 1 - \frac{\sigma_{\mathrm{off}_1}}{\sigma_{\mathrm{on}_1}}\frac{\sigma_{\mathrm{off}_2}}{\sigma_{\mathrm{on}_2}}
   \left(1+\frac{\langle I_{\mathrm{on}_1} \rangle}{\langle I_{\mathrm{off}_1} \rangle} + \frac{\langle I_{\mathrm{on}_2} \rangle}{\langle I_{\mathrm{off}_2} \rangle}\right)~.
\label{correction}
\end{equation}
The quantities $\langle I_{\mathrm{on}_1}\rangle$, $\langle I_{\mathrm{on}_2}\rangle$ represent
the mean ON-pulse intensities inside the selected pulse window over $N_\mathrm{p}$ pulses
for each of the bands, 1 and 2, after subtraction of the equivalent mean OFF-pulse
intensities, $\langle I_{\mathrm{off}_1}\rangle$, $\langle I_{\mathrm{off}_2}\rangle$.
The quantities, $\sigma_{\mathrm{on}_1}$, $\sigma_{\mathrm{on}_2}$, $\sigma_{\mathrm{off}_1}$ and
$\sigma_{\mathrm{off}_2}$ are the rms deviations of  $I_{\mathrm{on}_1}$, $I_{\mathrm{on}_2}$, $I_{\mathrm{off}_1}$,
and $I_{\mathrm{off}_2}$, respectively, again averaged over $N_\mathrm{p}$ pulses.
This expression corresponds to equation~(20) in~\citet{rickett1975}.

\begin{figure*}[p]

\includegraphics[height=5.6cm,width=9cm]{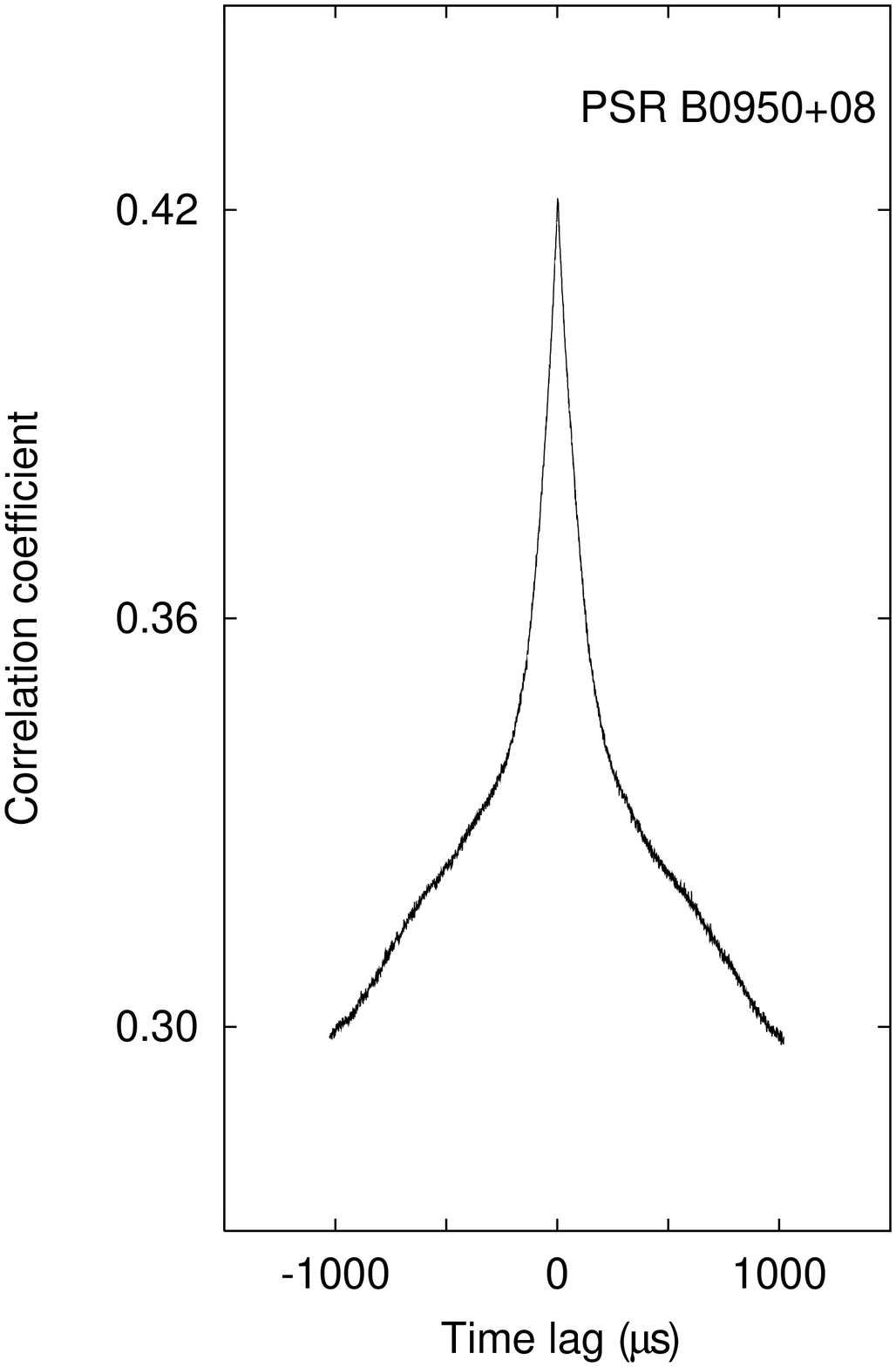}
\includegraphics[height=5.6cm,width=9cm]{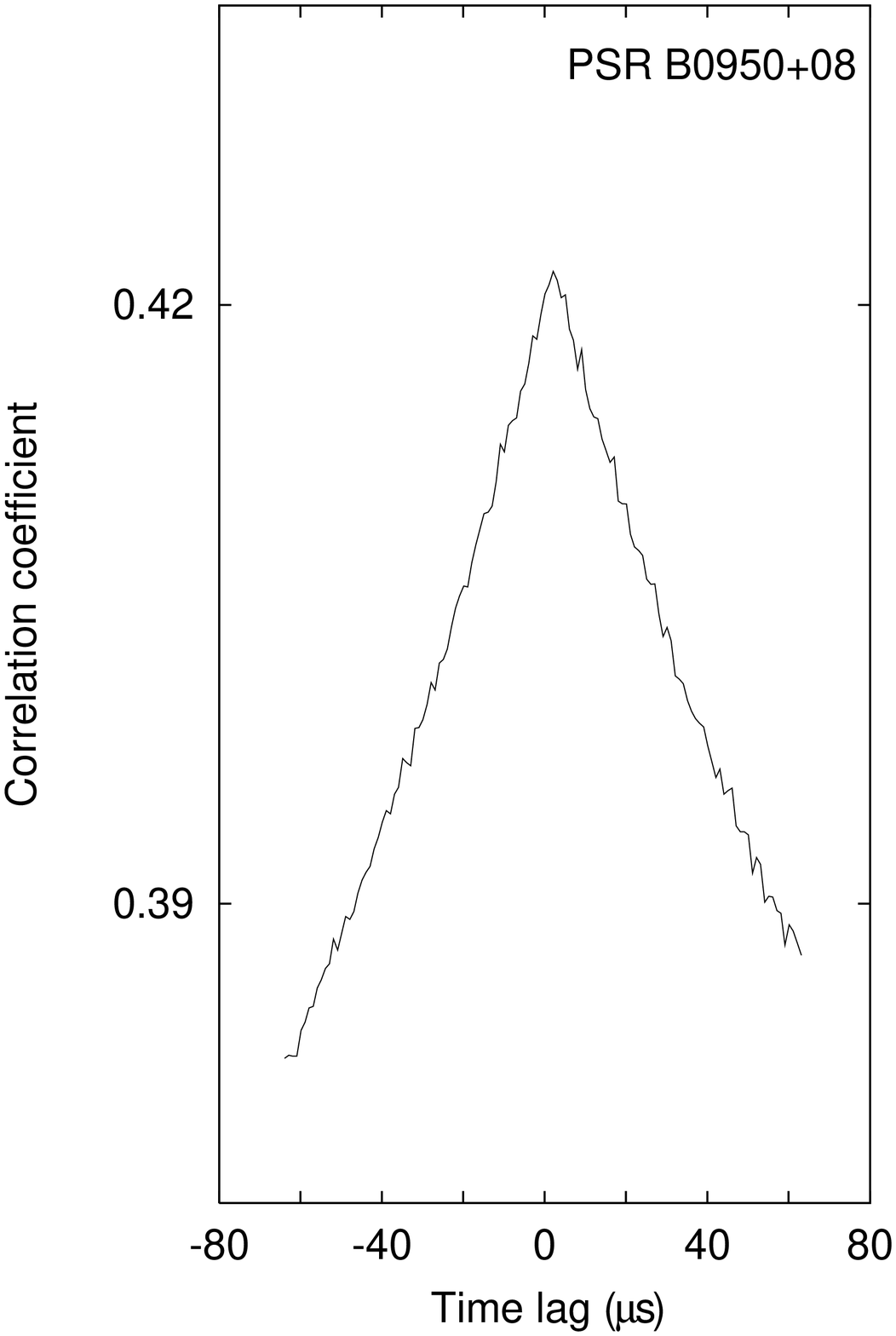}

\includegraphics[height=5.6cm,width=9cm]{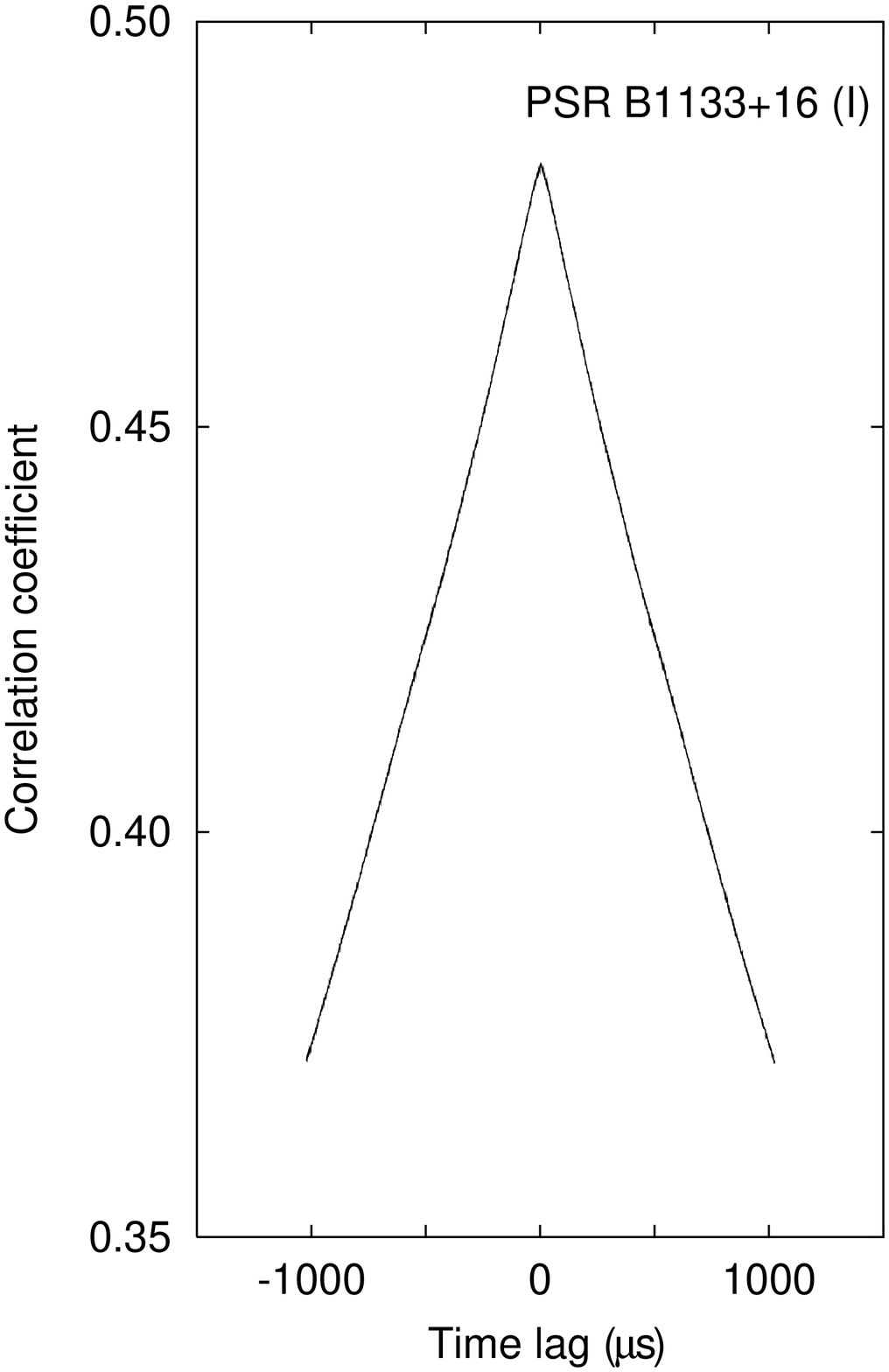}
\includegraphics[height=5.6cm,width=9cm]{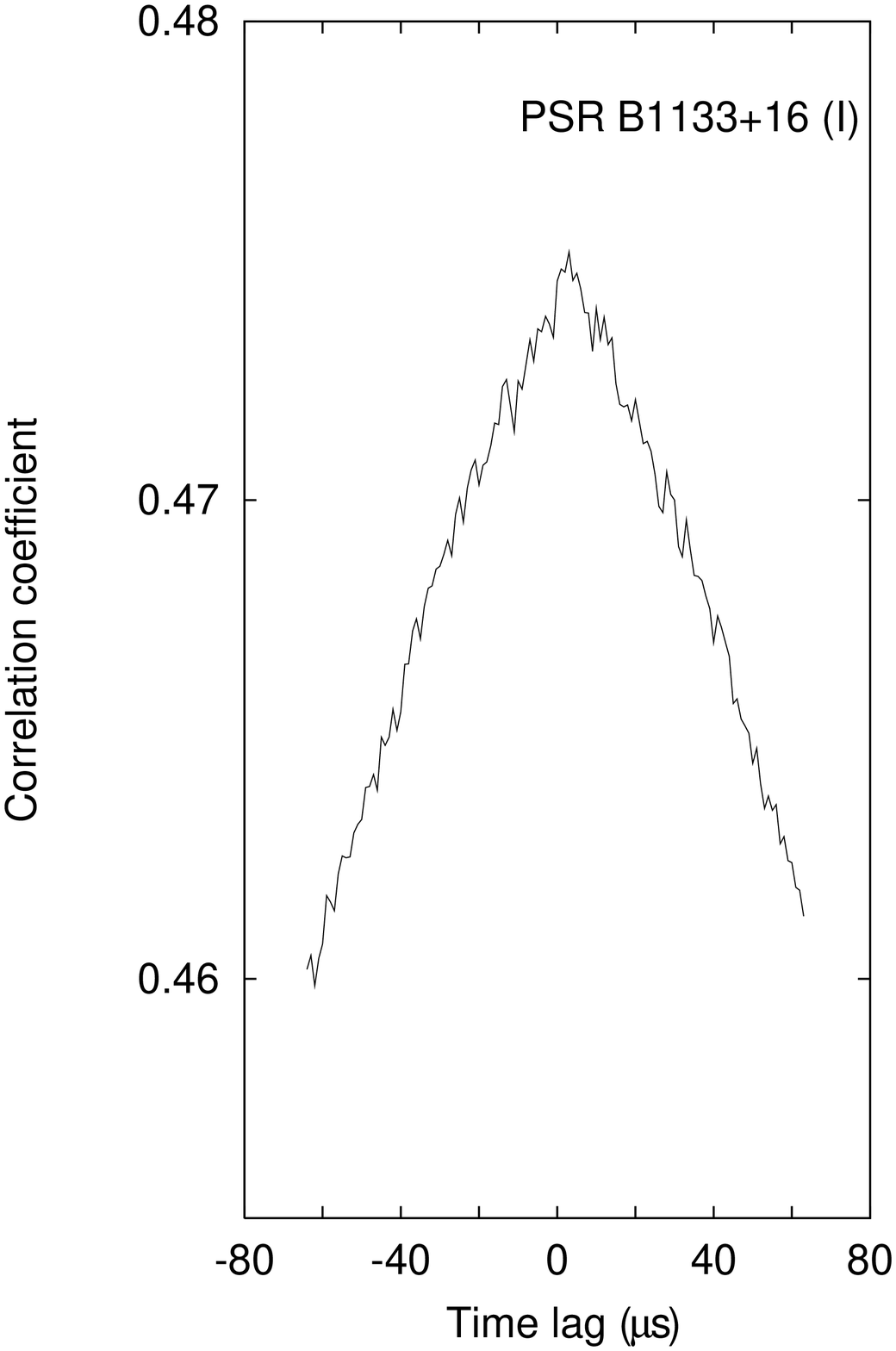}

\includegraphics[height=5.6cm,width=9cm]{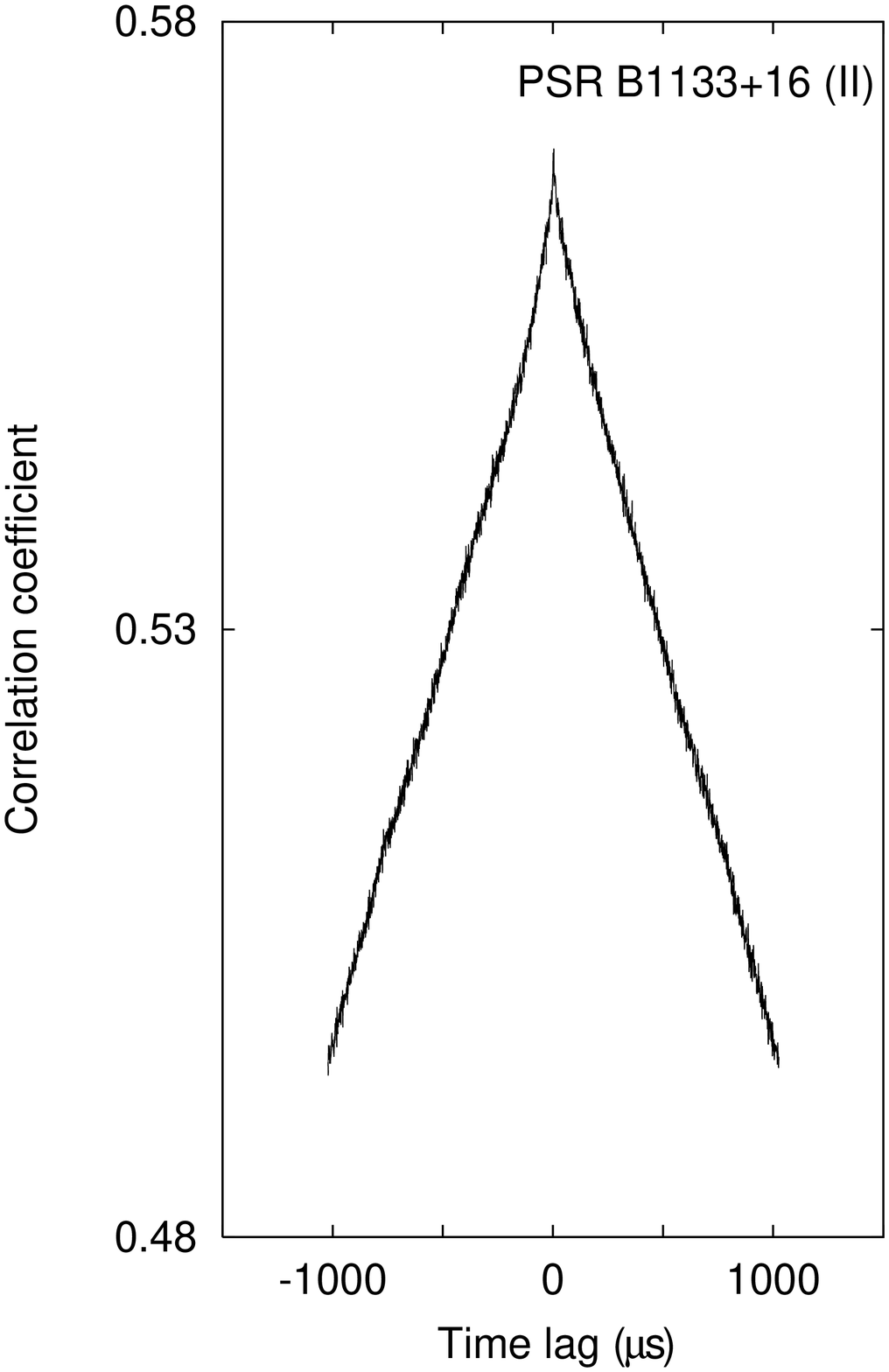}
\includegraphics[height=5.6cm,width=9cm]{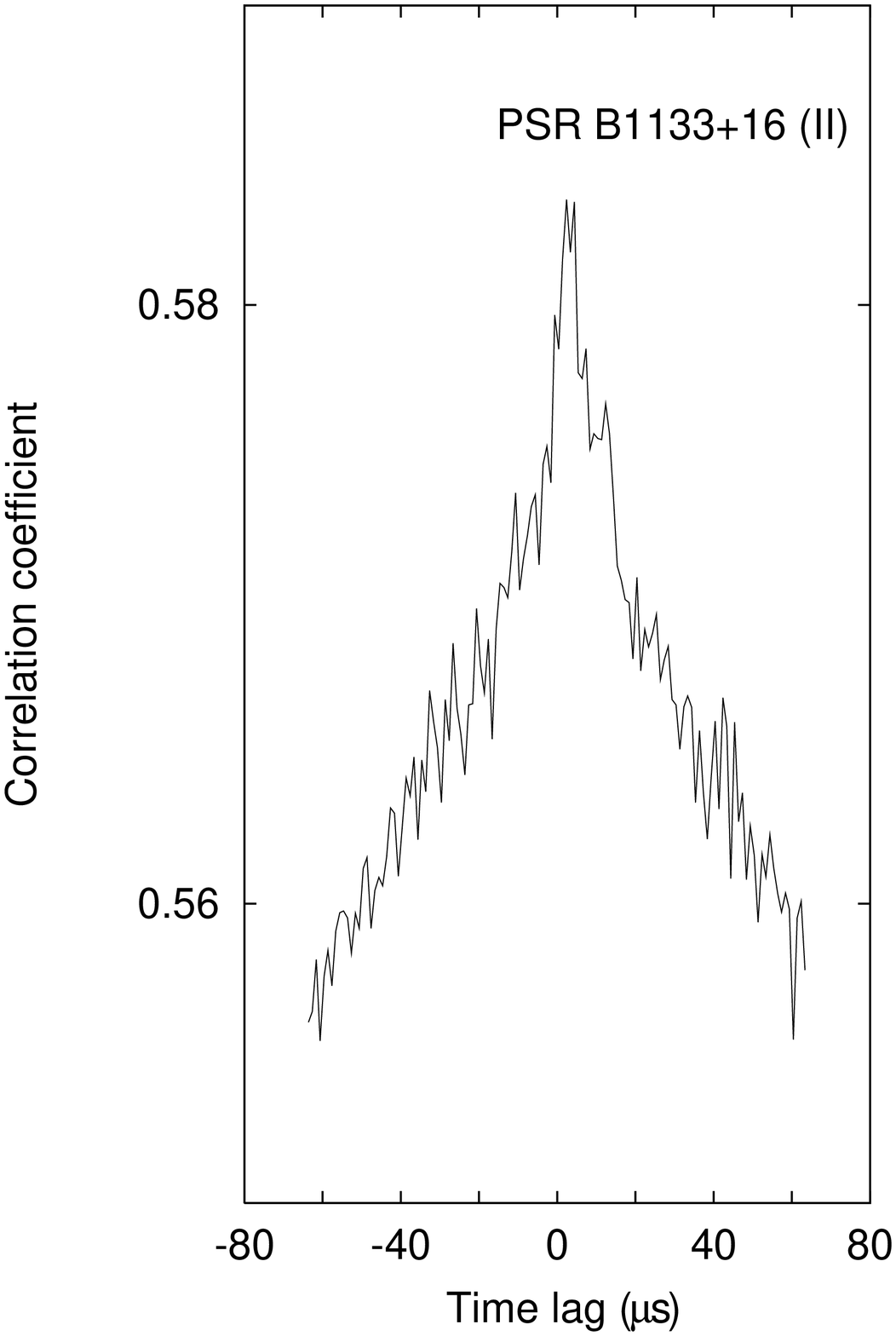}

\includegraphics[height=5.6cm,width=9cm]{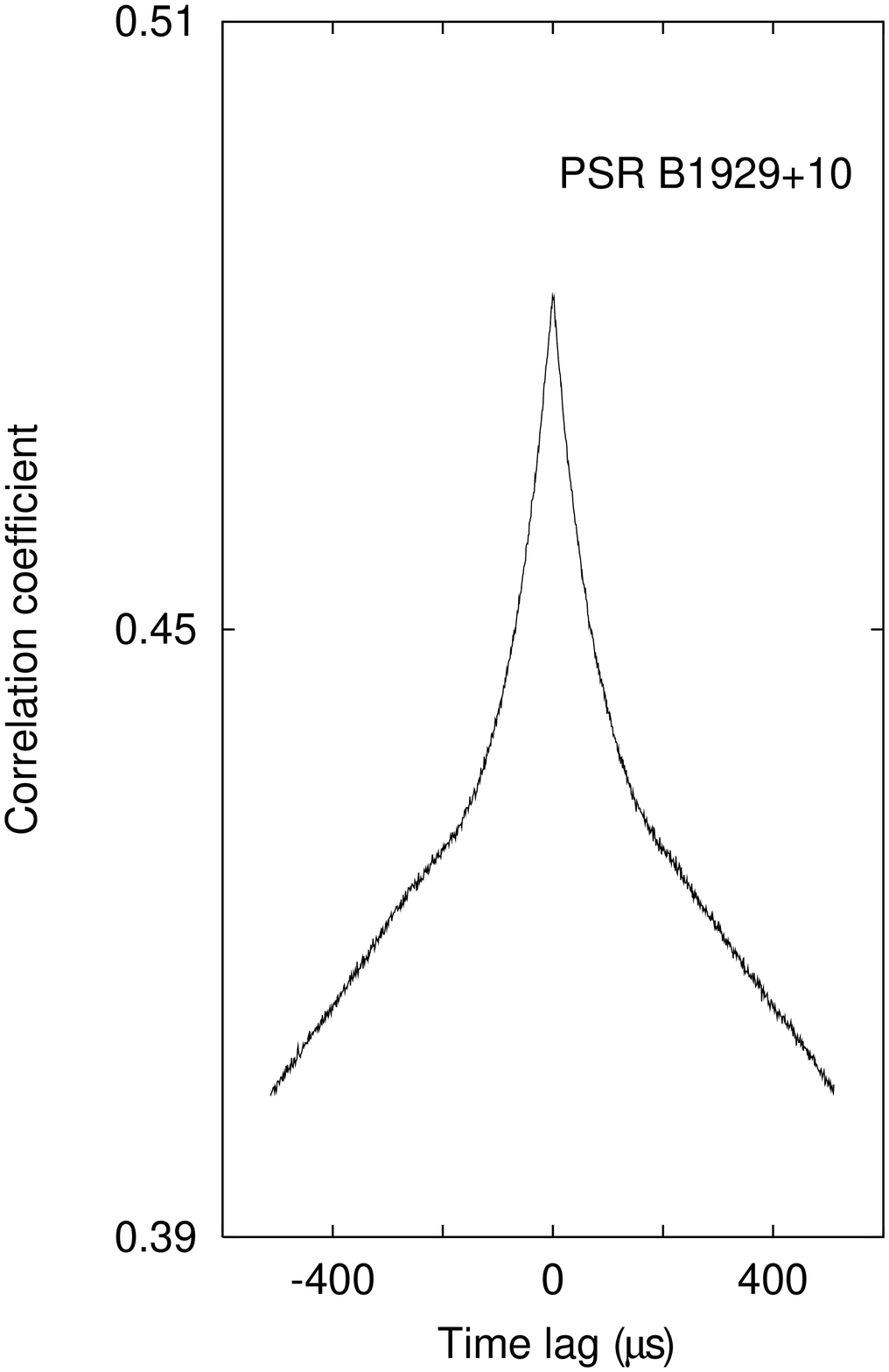}
\includegraphics[height=5.6cm,width=9cm]{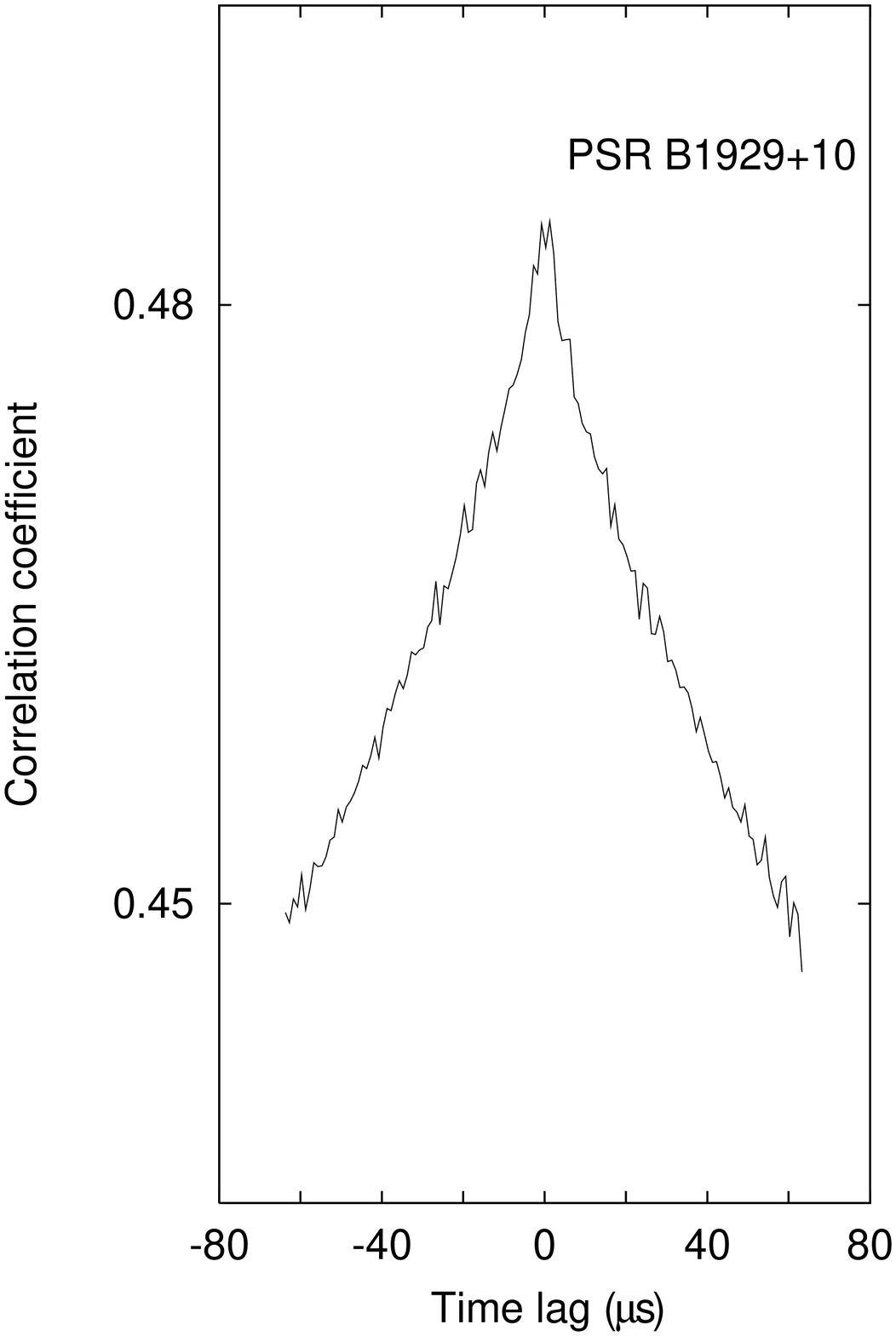}

\vskip 2mm

\caption{The average cross-correlation function (CCF) for PSR \object{B0950+08},
the first (I) and second (II) component of PSR \object{B1133+16},
and PSR \object{B1929+10}. The CCFs were calculated for the
unsmoothed ON-pulse intensities recorded in the conjugate 16-MHz bands and then
smoothed over a time interval of $1~\mic$. The right column of the plots represents the very
central portion of the CCFs shown in the left column.}

\label{avccf}

\end{figure*}

Figure~\ref{avccf} shows the CCFs smoothed to a time-lag resolution of $1~\mic$
and shifted so that their time-lag origins correspond to the delays $\delta
t_\mathrm{cal}$, computed from the dispersion measure, DM~
(see Table~\ref{psrpar1}). The left column shows the CCFs on a relatively
large time-lag scale while the right column displays only the central part of the CCFs.
The maxima of the CCFs are close to the value of 0.5 predicted by the
Amplitude-Modulated Noise (AMN) model~\citep{rickett1975}.
The relatively small discrepancies may be due to small and possibly insignificant
errors of the correction factors for the CCFs.

For PSRs \object{B0950+08} and \object{B1929+10} a sharp spike can be clearly
seen in the CCFs in the left column, indicating
correlated microstructure in the two bands with a well defined characteristic
timescale.
For \object{PSR~B1133+16} a central spike is less pronounced in either
of the two components at this frequency, but also existent.

We determined the characteristic timescale of the typical width of the microstructure
from the intersection of the linear slope of the central narrow part (excluding
the most inner section, see below) and the more slowly decreasing
broader part of the CCFs by linear least-squares fits.
We list the values with the corresponding statistical
standard errors as $\tau_\mathrm{\mu-broad}$  in Table~\ref{psrpar2}.

Our value for $\tau_\mathrm{\mu-broad}$ for \object{PSR~B0950+08} is
compatible with earlier measurements between 130 and
$200~\mic$~\citep{rickett1975, cordes1977, hankins1978, lange1998}. For
\object{PSR~B1133+16}  microstructure was reported in the range of $\sim 340$ to
$\sim 650~\mic$~\citep{hankins1972, ferguson1976, ferguson1978, cordes1976a, popov1987, lange1998},
comparable to our value of component I but three to sixfold larger than that for component II.
For \object{PSR~B1929+10} the width of microstructure has never been measured in the average ACF
or CCF before.
\begin{figure*}[t]

\includegraphics[width=5.5cm,height=6cm]{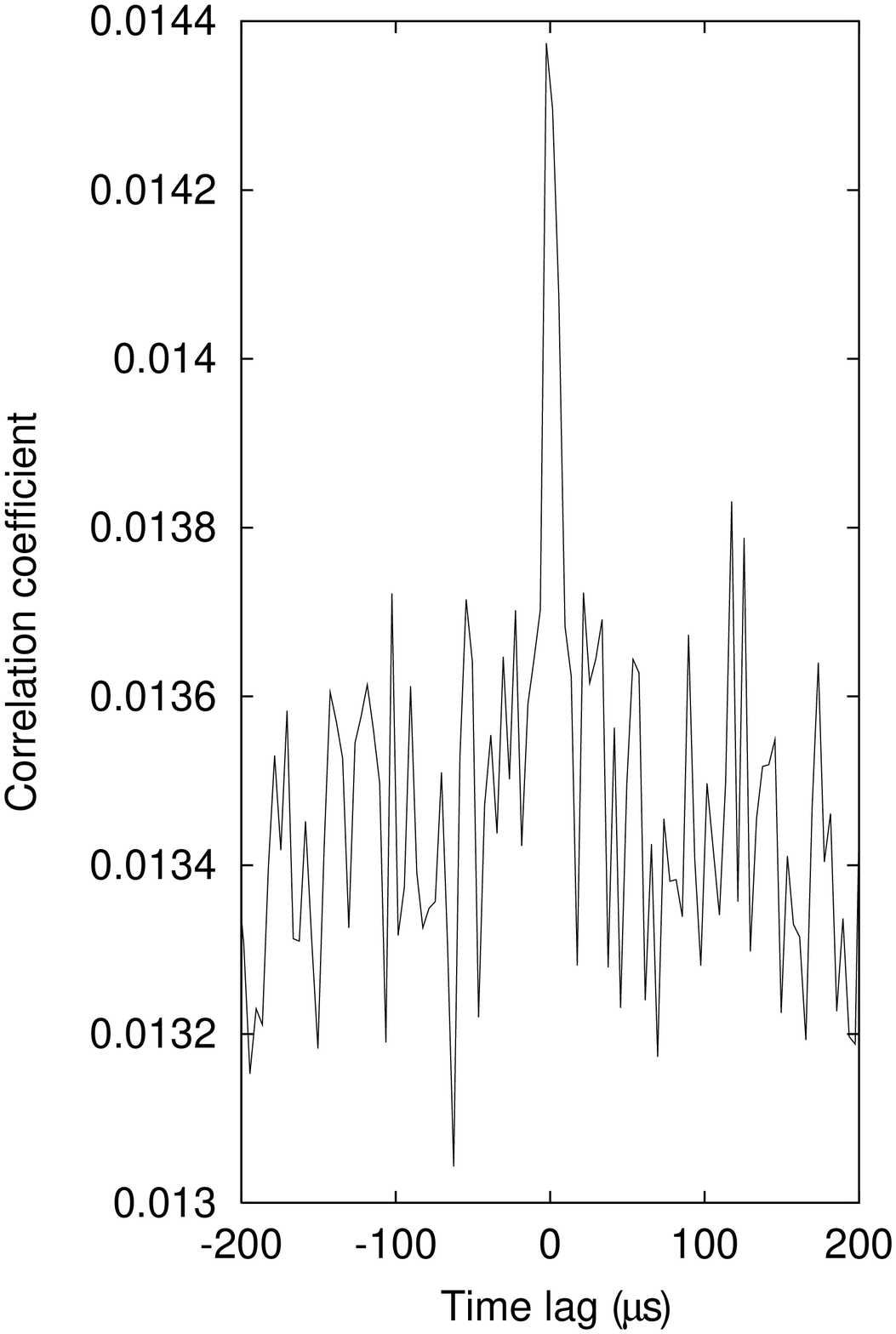}
\includegraphics[width=5.5cm,height=6cm]{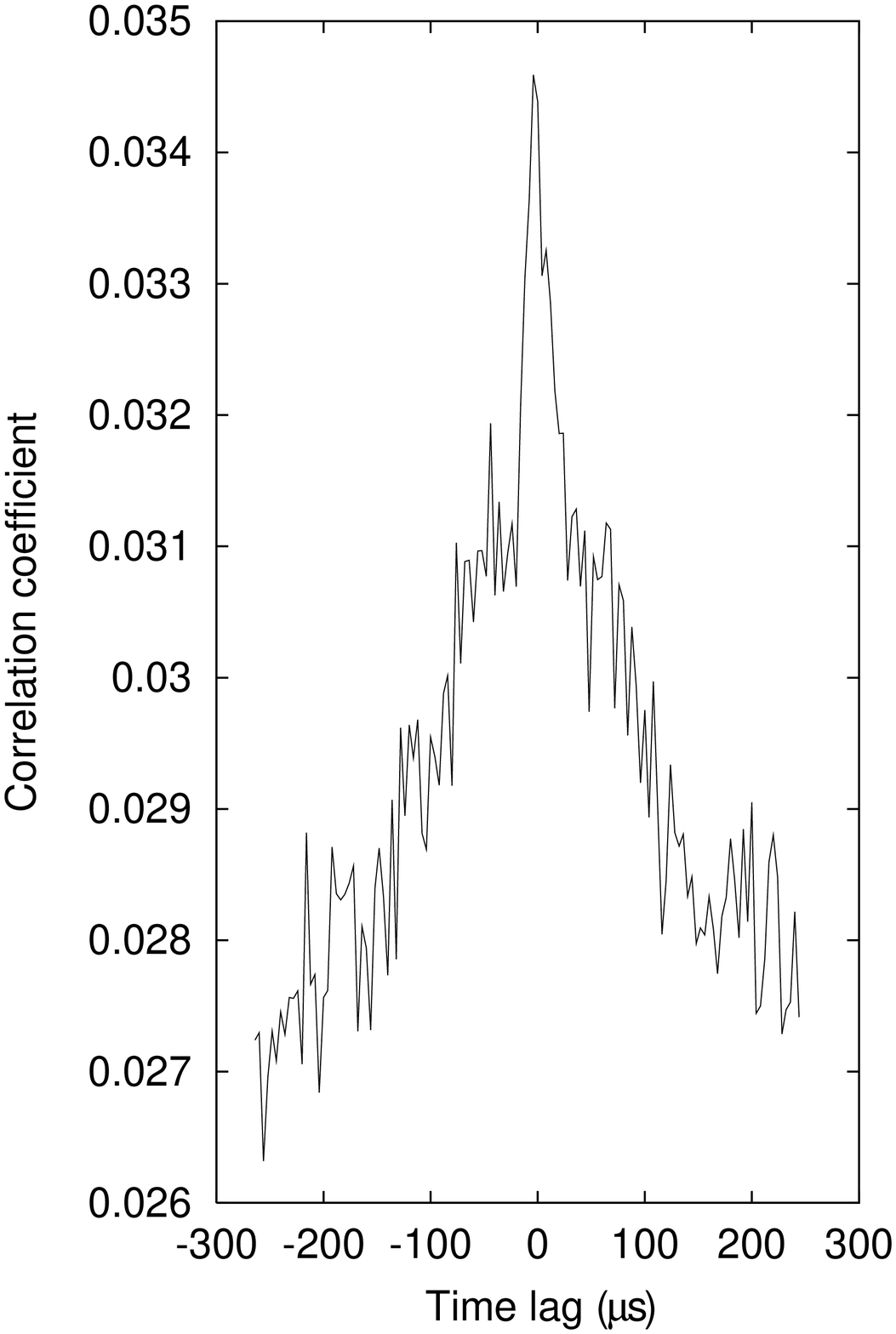}
\includegraphics[width=5.5cm,height=6cm]{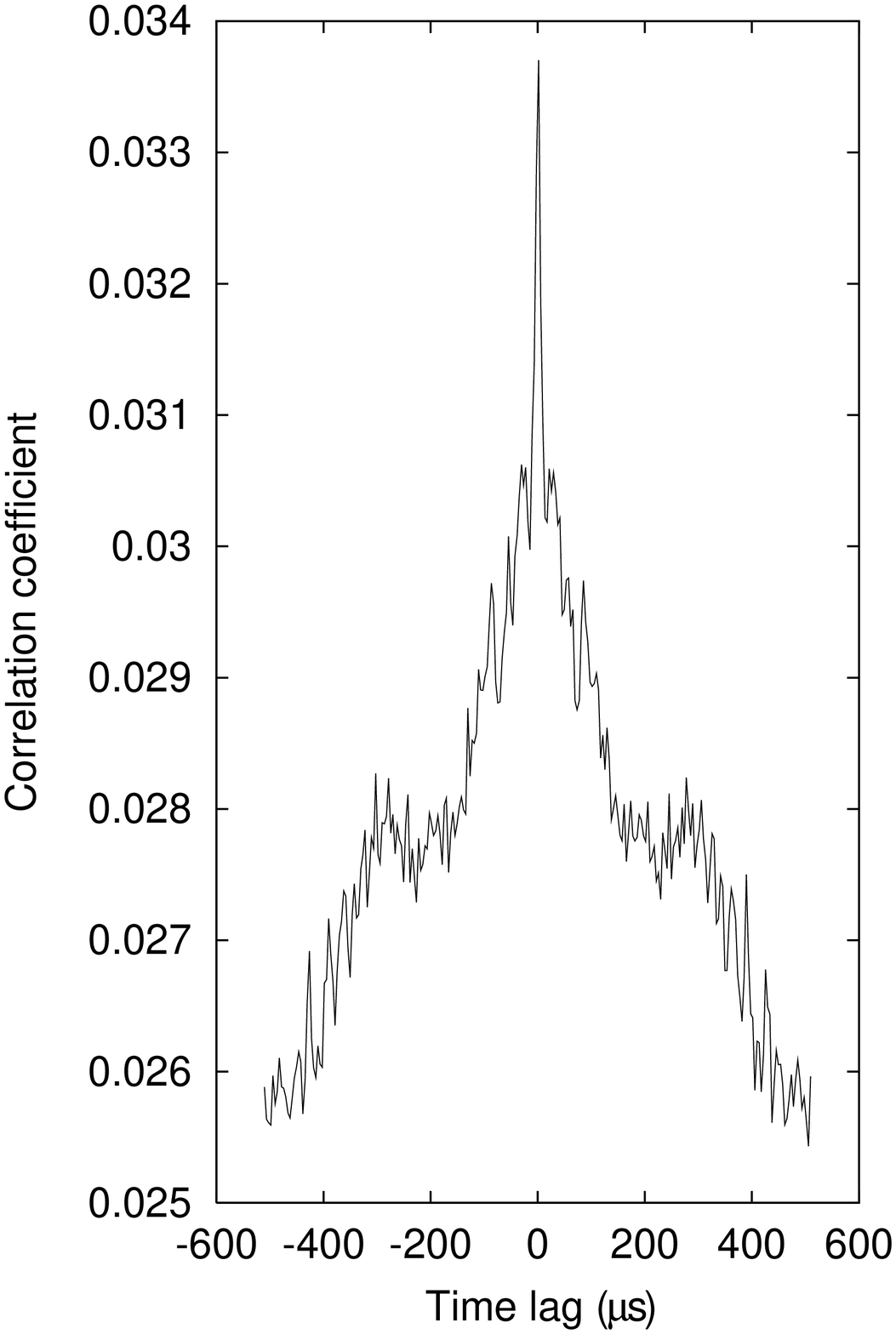}

\vskip 2mm

\caption{Examples of individual CCFs for \object{PSR~B1133+16} (component II)
in order of increasing complexity. The left plot shows microstructure with one
notable width only. The middle and the right plots show microstructure
with two and three widths. The scale of the correlation coefficients is not
corrected for receiver noise.}

\label{exccf}

\end{figure*}
However,~\citet{lange1998} reported on the detection of
structure on timescales around $150~\mic$ close to their effective time
resolution of $70~\mic$.

In the  right column we display the central portions of the CCFs.
For \object{PSR~B1133+16} (I) no additional abrupt sharpening of the
CCFs can be seen. However for PSRs \object{B0950+08}, \object{B1133+16} (II),
and \object{B1929+10} an additional central spike was found,
indicating particularly short microstructure with characteristic timescales,
$\tau_\mathrm{\mu-narrow}$, of order $10~\mic$ (see Table~\ref{psrpar2}).
Such short microstructure  has never been seen in the
average  ACF or CCF of any pulsar. However, for the giant pulses from the
Crab pulsar, microstructure was found in the ACF of a single pulse
with the width at the 25\% level of the ACF of $\leq
1~\mic$~\citep{hankins2000}.

\subsubsection {Time delay of microstructure from the average cross-correlation
function}
\label{time-delay}

\noindent
A close inspection of the inner part of the CCFs shows that the peak does not
always occur exactly at the expected time delay. To quantify the discrepancy
we measured the center of symmetry of the top part of the central portion of
the CCFs in Figure~\ref{avccf} and list it as $\delta t_\mathrm{obs}$ in Table~\ref{psrpar2}.
To determine the standard errors, $\Delta\delta t_\mathrm{obs}$,
we followed~\citet{chashei1975} and used

\begin{center}
\begin{equation}
\Delta\delta t_\mathrm{obs} \sim \frac
{4\tau_\mathrm{\mu}}{\mathit{SNR}}(N_\mathrm{p}T)^{-1/2}~,
\label{eqthird}
\end{equation}
\end{center}
\noindent
where $\tau_\mathrm{\mu}$ is the microstructure timescale, $\tau_\mathrm{\mu-broad}$,
listed in Table~\ref{psrpar2},~
and {\em SNR} is the signal-to-noise ratio of the intensity of the average single
pulse integrated  over the pulse window (Table~\ref{psrpar1}).

For \object{PSR~B1929+10} the observed time delay is equal within
$\Delta\delta t_\mathrm{obs}$ to the expected
time delay. However, for \object{PSR~B0950+08} and the first and second
component of \object{PSR~B1133+16} there are significant discrepancies. The differences,
$\delta t_\mathrm{obs} - \delta t_\mathrm{cal}$, are about $-2~\mic$
(see Table~\ref{psrpar2})
in the sense that micropulses at the higher frequency arrive slightly later than
expected in comparison to the micropulses at the lower frequency.
If interpreted as a dispersion measure difference,
they  correspond to values of DM, 1 to 2\% smaller than those
listed in Table~\ref{psrpar1}. However, as we will discuss in
Section~\ref{discuss}, we do not think that the values of DM in
Table~\ref{psrpar1} need to be corrected.

\begin{figure*}[t]

\hskip 15mm\includegraphics[width=6cm,height=6cm]{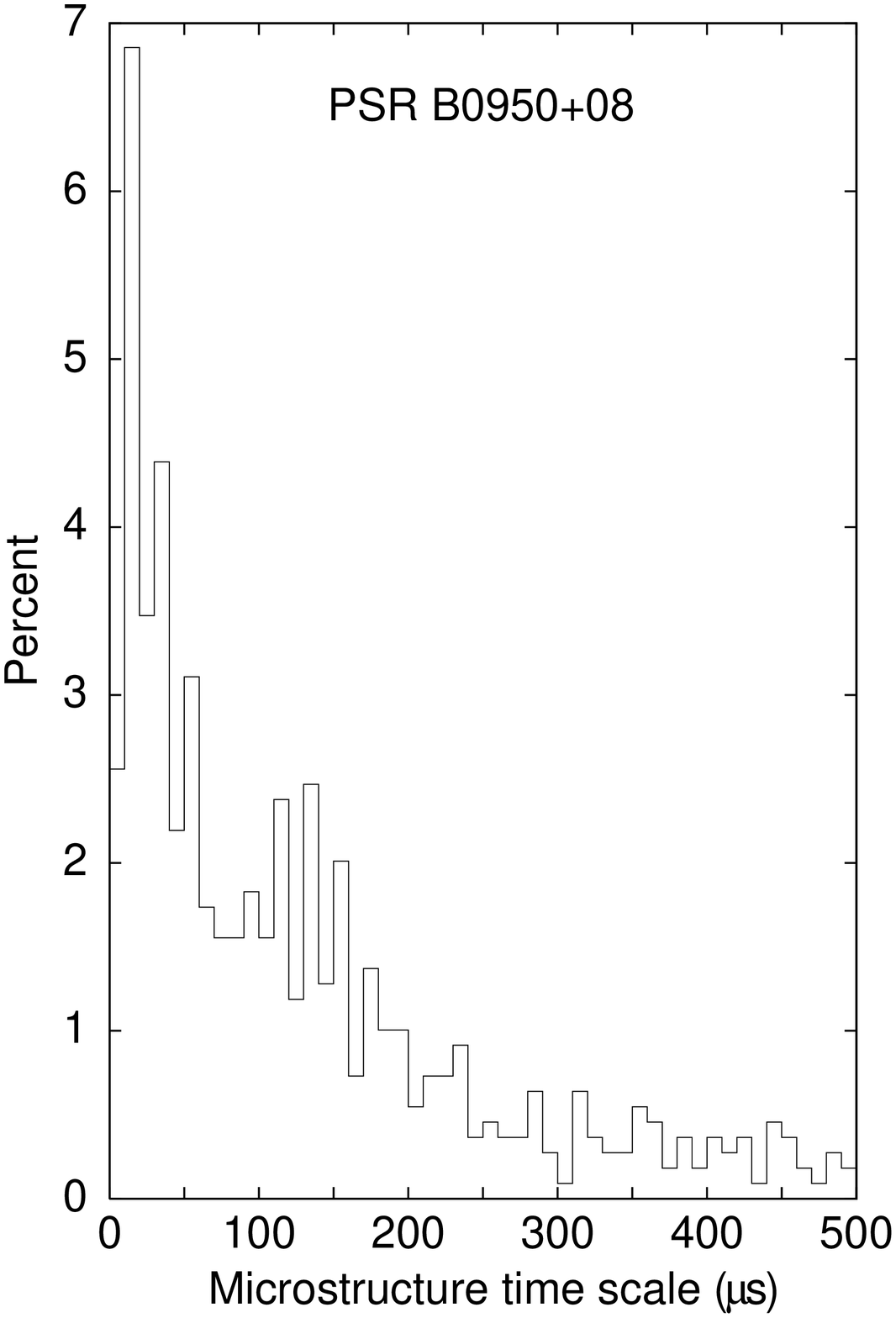}
\hskip 10mm\includegraphics[width=6cm,height=6cm]{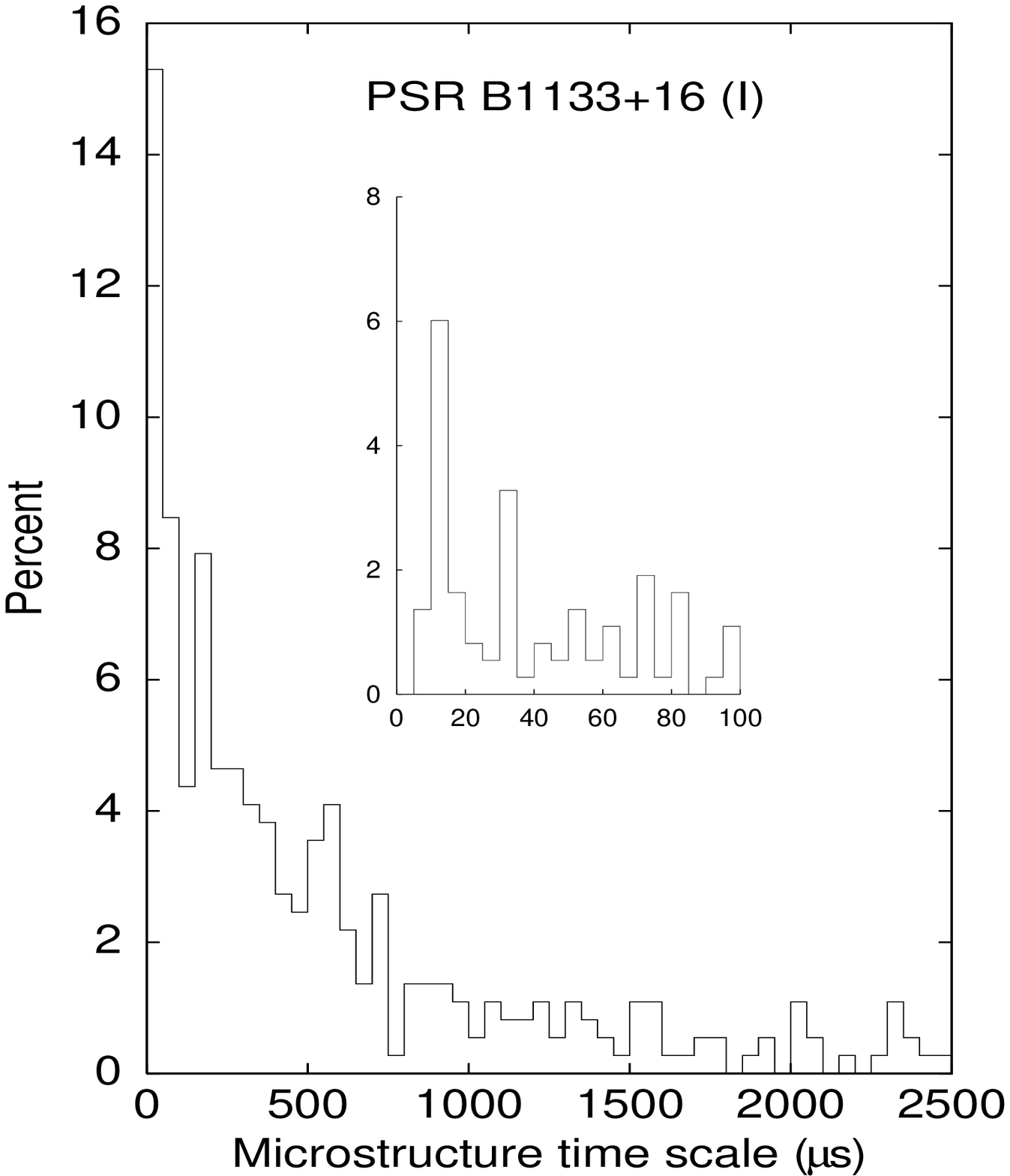}

\hskip 15mm\includegraphics[width=6cm,height=6cm]{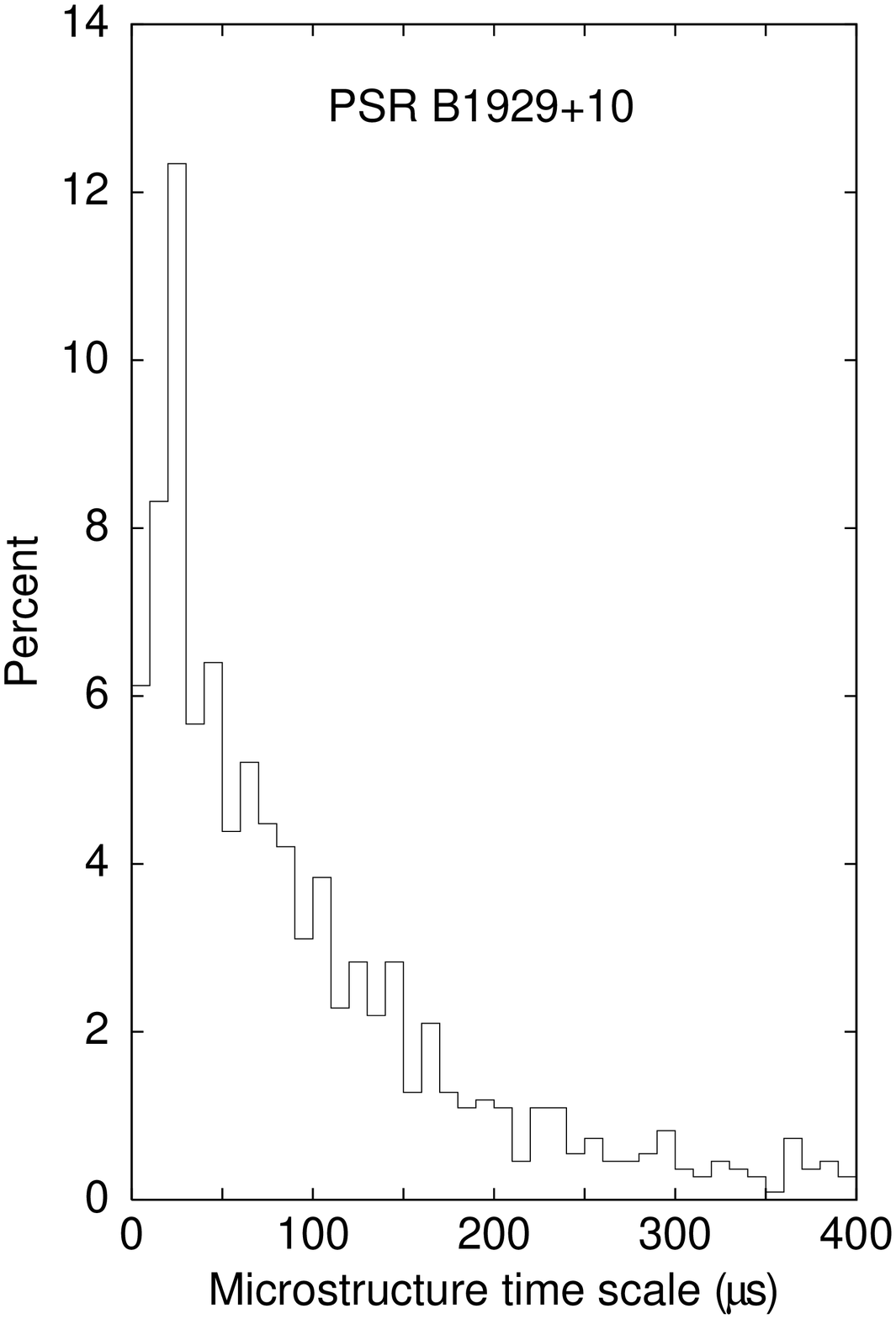}
\hskip 10mm\includegraphics[width=6cm,height=6cm]{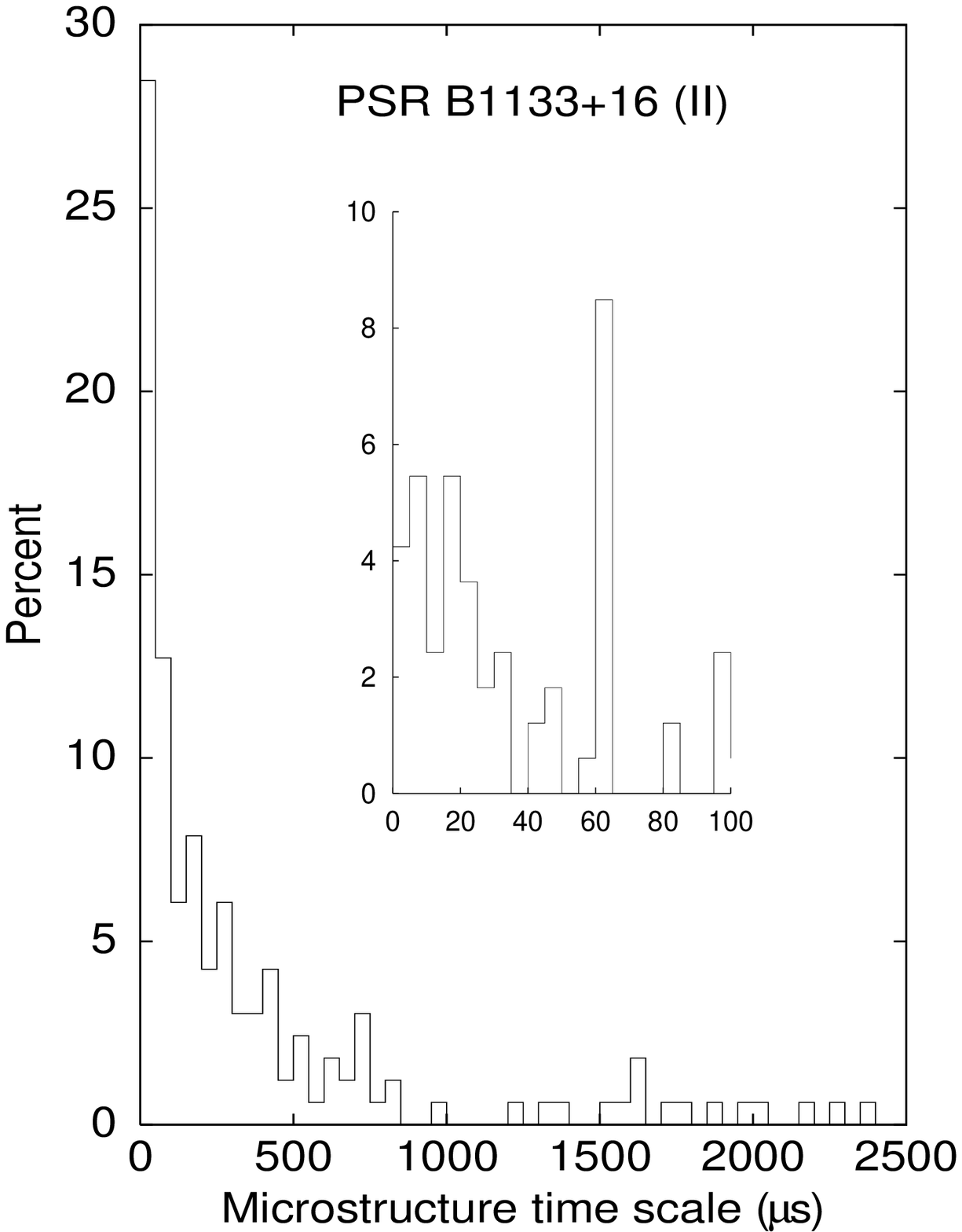}

\vskip 2mm

\caption{Histograms of timescales of detected microstructure in the individual
CCFs of single pulses for PSRs \object{B0950+08}, \object{B1133+16} (components I, II),
and \object{B1929+10}. The insets display histograms for the
shortest timescales.}

\label{histtau}

\end{figure*}

\subsubsection {The distribution of microstructure widths from individual
cross-correlation functions} \label{distrib-tau-mu}

\noindent
In order to determine the distribution of microstructure widths for individual
pulses, we computed the CCF for each of the $N_\mathrm{p}$ selected strong
single pulses. Note that for \object{PSR~B1929+10} the number in
Table~\ref{psrpar1} is given in parentheses, different from the number of pulses
used to compute the average CCF. An interactive ``TV''-based task was used
to display the individual CCFs in a variety of timescales. With the help of
a TV-cursor, the microstructure width was determined, again, from
that point of the CCF where the central steep slope flattens out.

The majority of the selected strong pulses showed distinct short-term
microstructure in the individual CCFs.
Three examples of individual CCFs are shown in Figure~\ref{exccf}
for \object{PSR~B1133+16} (component II). The left
plot shows a CCF with an individual spike only, the middle  and the
right plots show more complex CCFs with two and three different widths for the
microstructure. For PSRs \object{B0950+08} and \object{B1929+10} very often several,
up to five, different structural scales were found in the same pulse.
In these cases several values for the microstructure width  were obtained from
one pulse. The total number of values for the microstructure width is listed
for each pulsar (and each component in case of \object{PSR~B1133+16})
in Table~\ref{psrpar1} as $N_\mathrm{\mu}$.
A number of single pulses had smooth CCFs without any distinct microstructure
features. That number is given as $N_\mathrm{no-\mu}$ in Table~\ref{psrpar1}.
Only one  from 225 pulses of  \object{PSR~B0950+08}, or less than 0.5\% had no
microstructure. For \object{PSR~B1929+10} about 3\% of single pulse had no microstructure,
while for  \object{PSR~B1133+16} about 17\% and 38\% of pulses
had no microstructure in case of the first and the second
component, respectively. This finding indicates even higher percentages of
single pulses with microstructure than reported
earlier~\citep[e.g.,][]{smirnova1994, lange1998}, if the results also hold for the
weaker pulses which had to be ignored in our analysis because of sensitivity
reasons.

The histograms of the microstructure widths are presented in
Figure~\ref{histtau}. For all three pulsars they are skewed towards shorter widths.
In general, they show a moderate rise from broader to shorter width starting at
$\sim 200~\mic$ for PSRs \object{B0950+08} and \object{B1929+10} and $\sim
800~\mic$ for the two components of \object{PSR~B1133+16}.
The rise becomes markedly sharper at $\sim 40~\mic$ for PSRs \object{B0950+08}
and \object{B1929+10} and at $\sim 200~\mic$
for each of the components of \object{PSR~B1133+16}, or at $\sim 0.08$ and $\sim 0.02\%$
of P, respectively.
The histograms peak at
$\sim 20$ to $30~\mic$ for PSRs \object{B0950+08} and \object{B1929+10}
and $\leq 50~\mic$ for the two components of \object{PSR~B1133+16}.
Towards shorter widths the histograms of PSRs \object{B0950+08},
\object{B1133+16} (I, see inset of plot), and \object{B1929+10} display
a sharp cutoff at a width of $\sim10~\mic$, and of \object{PSR~B1133+16}~(II) at $< 10~\mic$
(not visible in plot).
The shortest widths measured are $7~\mic$ for \object{PSR~B0950+08},
$6~\mic$ and $2~\mic$ for \object{PSR~B1133+16}
(component I and II, respectively), and $5~\mic$ for \object{PSR~B1929+10}
(Table~\ref{psrpar2}).
The cutoff width and the shortest widths measured are greater than a) the
smoothing interval of $1~\mic$ used in the analysis of the individual CCFs
and b) the expected scattering time for either of the pulsars, and are therefore
intrinsic properties of the pulsars.
Again, there is a difference for the first and second component
of \object{PSR~B1133+16}. Micropulses with a width $\leq 50~\mic$ occur almost
twice as frequent in component II than in component I.

\subsection{Short-term noiselike intensity fluctuations}

\noindent
Our average CCFs as well as our individual CCFs do not show any microstructure
features with a sub-microsecond timescale.
On the other hand, as was mentioned in Section~\ref{ind_pulse},
strong pulses when plotted with the highest time resolution
contain  bright sub-microsecond fluctuations (see Figure~\ref{giant}).
Also,~\citet{sallmen1999} reported to have found intensity fluctuations from the
Crab pulsar which were still unresolved at their highest time resolution of 10~ns. Is it
possible that ``nanopulses'' exist in our data with a width not much
larger than our highest time resolution of 62.5~ns?
They may perhaps not be visible in our histograms with
more than 100 times wider bins. They could perhaps also be largely uncorrelated
for the two bands and therefore
not apparent in the CCFs. To investigate the significance of our fast intensity
fluctuations we have to compare their statistics with the statistics of noise.

We use two approaches in our data analysis: 1) computation
of short-term ACFs with a time-lag resolution of 62.5~ns for ON-pulse and OFF-
pulse windows, and 2) comparison of the distribution of the intensities
of short-term (62.5~ns) fluctuations ON-pulse with the distribution of
such intensities OFF-pulse and also with the $\chi^2$-distribution for thermal noise.

\subsubsection{Short-term ACFs}

\noindent
Short-term ACFs, $R(\tau)$, with

\begin{center}
\begin{equation}
R(\tau)=\frac{1}{R(0)}\sum_{t=1}^T [\,I(t)-\langle I\rangle \,][\,I(t+\tau)-\langle I\rangle \,]~,
\label{eqforth}
\end{equation}
\end{center}
\noindent
were computed for single pulses from many short adjacent
intervals of the data with detected signal, $I(t)$, at sample number,
$t$, both in ON-pulse and OFF-pulse windows. The short-time intervals
were each $8~\mic$ long and contained $T=256$ samples.
The instantaneous mean level $\langle I \rangle$ used for the subtraction is given as
$\langle I \rangle=\sum_{t=1}^TI(t)/T$. For each single pulse
$m=256$ ACFs were computed for the ON-pulse window
covering an interval of 2.048 ms  centered close to the maximum of the selected
pulses.
The same number of ACFs was also computed for the OFF-pulse window of each
single pulse. Note, that the definition of the ACF used here is not quite
equivalent to the definition of the CCF in the equation~(\ref{eqsec}).

In order to obtain sufficient sensitivity, we selected only very strong single
pulses of number $k$: 41 for \object{PSR~B0950+08}, 24 for \object{PSR~B1133+16}, and 30
for \object{PSR~B1929+10}. For each of the pulsars  no systematic
difference was apparent between the ON-pulse and OFF-pulse average ACFs above
the
$3\sigma_\mathrm{ACF}$ level. For an
average ACF of white noise: $\sigma_\mathrm{ACF}=\frac{1}{\sqrt{Tmk}}$, or $\sim 0.001$ in our cases.
The corrections to the ON-pulse ACFs for receiver noise
(see equation~(\ref{correction})) are only $5-10\%$.
Therefore, since no deviations were found between ON-pulse and OFF-pulse ACFs
above the level of 0.003, we conclude that submicrosecond micropulses, if
present at all, do not contribute to the total power of pulsar signal variations by more than
$0.3~\%$.

\subsubsection{Distribution of intensities}

\begin{figure}[htb]

\hskip 5mm\includegraphics[width=8cm,height=10.5cm]{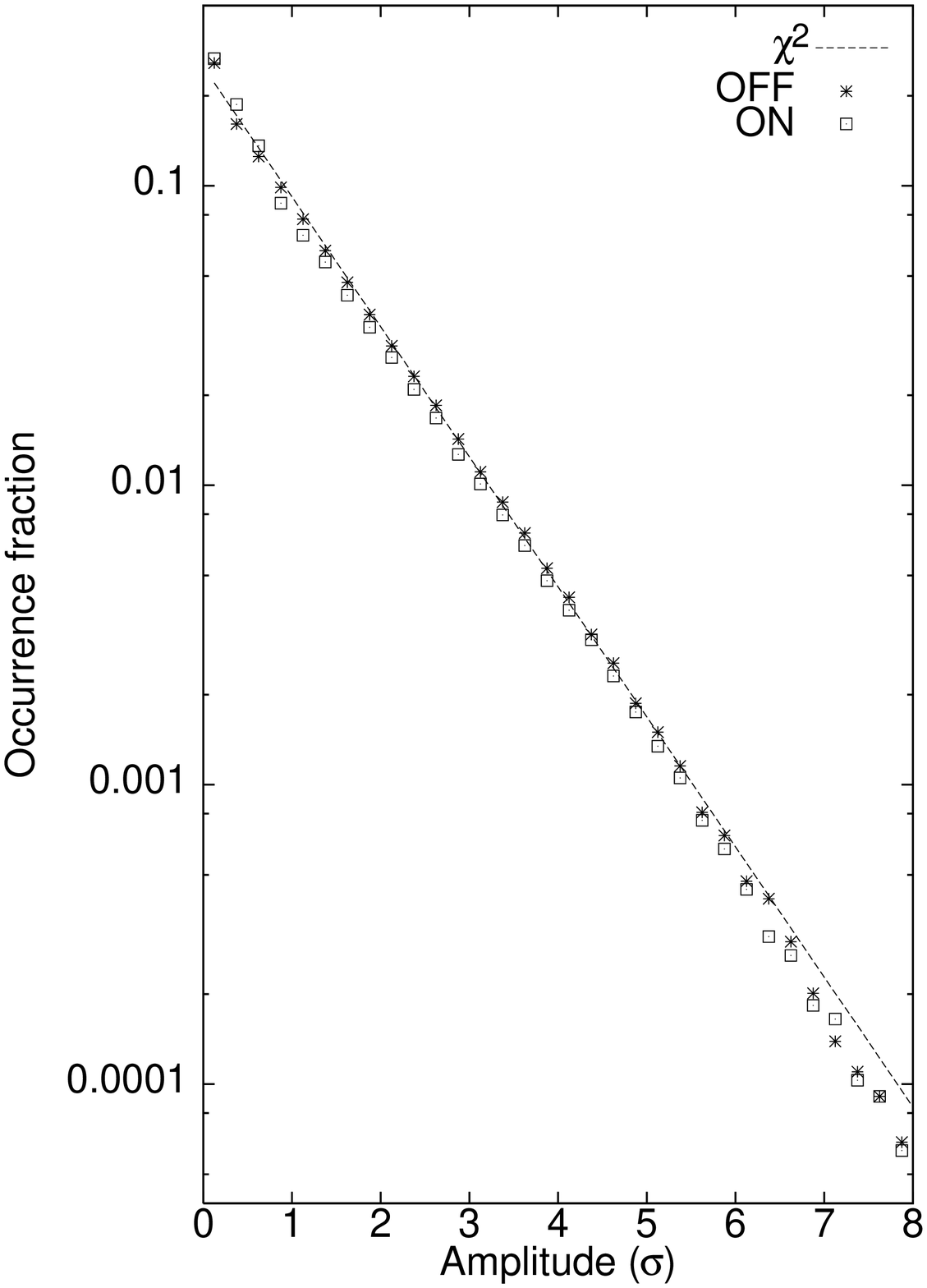}

\vskip 2mm

\caption{Histograms of intensities at the highest resolution of 62.5~ns
recorded ON-pulse (rectangles) and OFF-pulse  (stars) for three extremely strong
single pulses of \object{PSR~B1133+16} (component I). The dotted line
corresponds to the $\chi^2$-distribution for thermal noise.
The ordinate gives the number of occurrences
as a fraction of the total number of intensity samples of 393\,216. The abscissa
gives the intensity in units of~$\sigma$.}

\label{dist}

\end{figure}

\noindent
We also  compared the ON-pulse with the OFF-pulse intensity
distributions and the $\chi^2$-distribution for thermal noise for the data with the
highest resolution of 62.5~ns. For each
8-$\mic$ interval with $T=256$ samples, the rms deviation $\sigma$ and
 the quantity, $(I(t) - \langle I \rangle)/\sigma$, were computed.
  The results obtained for three extremely strong pulses of
\object{PSR~B1133+16} (component I) are presented in Figure~\ref{dist}.
Both distributions are fairly well fit by the
theoretical curve for the $\chi^2$-distribution. 
The deviations at large amplitudes of both the ON-pulse and OFF-pulse 
distributions from the theoretical line
can be contributed to effects of the two-bit sampling. In general, no notable 
differences are apparent between the ON-pulse
and OFF-pulse distributions.
ON-pulse and OFF-pulse intensity
fluctuations much shorter than $8~\mic$ therefore have the statistics of thermal
noise, consistent with the AMN model~\citep{rickett1975}.
The bright unresolved intensity spikes displayed in Figure~\ref{giant} are
therefore insignificant statistical outbursts. No nanopulses were found.

\begin{figure*}[t]

\includegraphics[width=7cm,height=5cm]{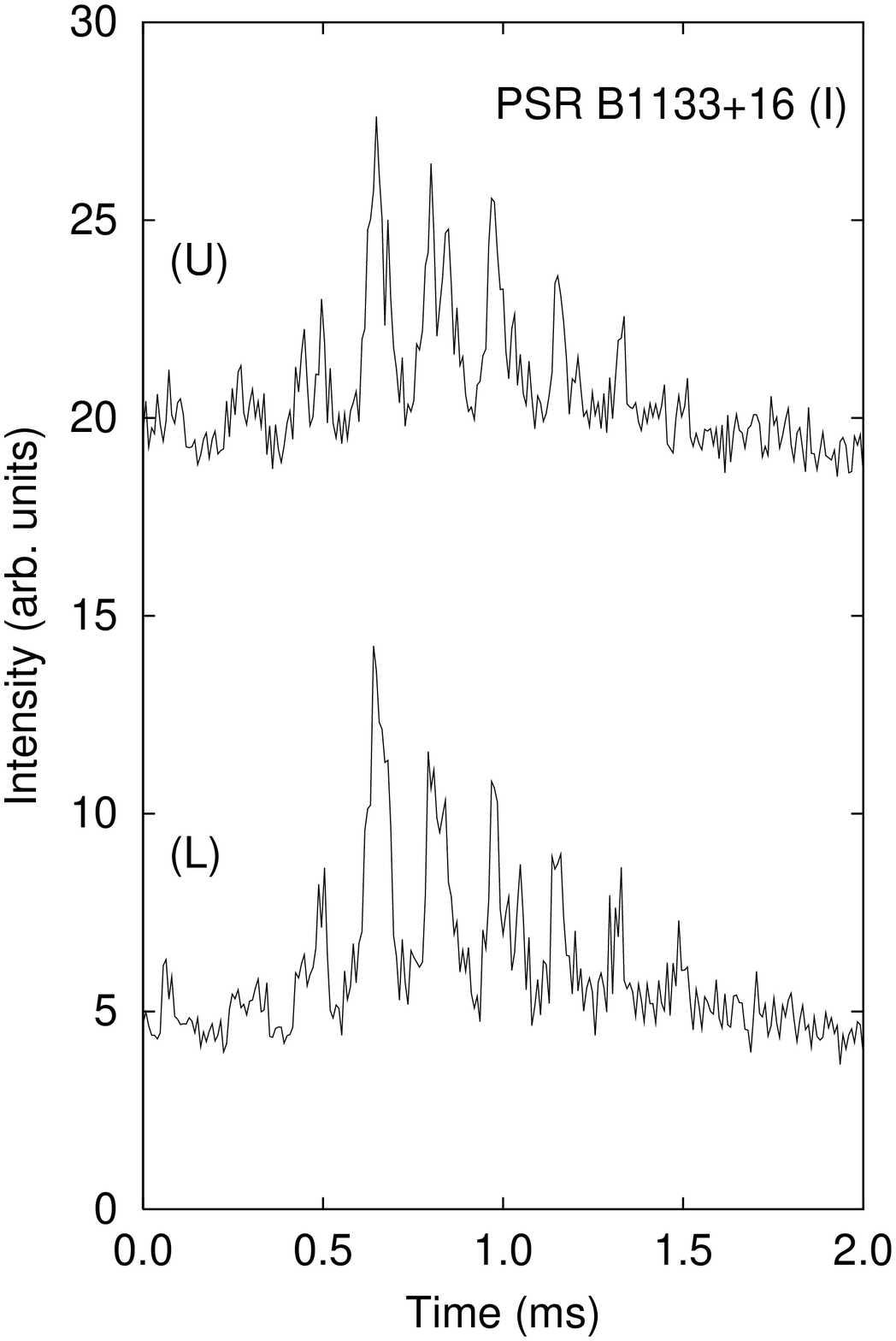}
\includegraphics[width=5cm,height=5cm]{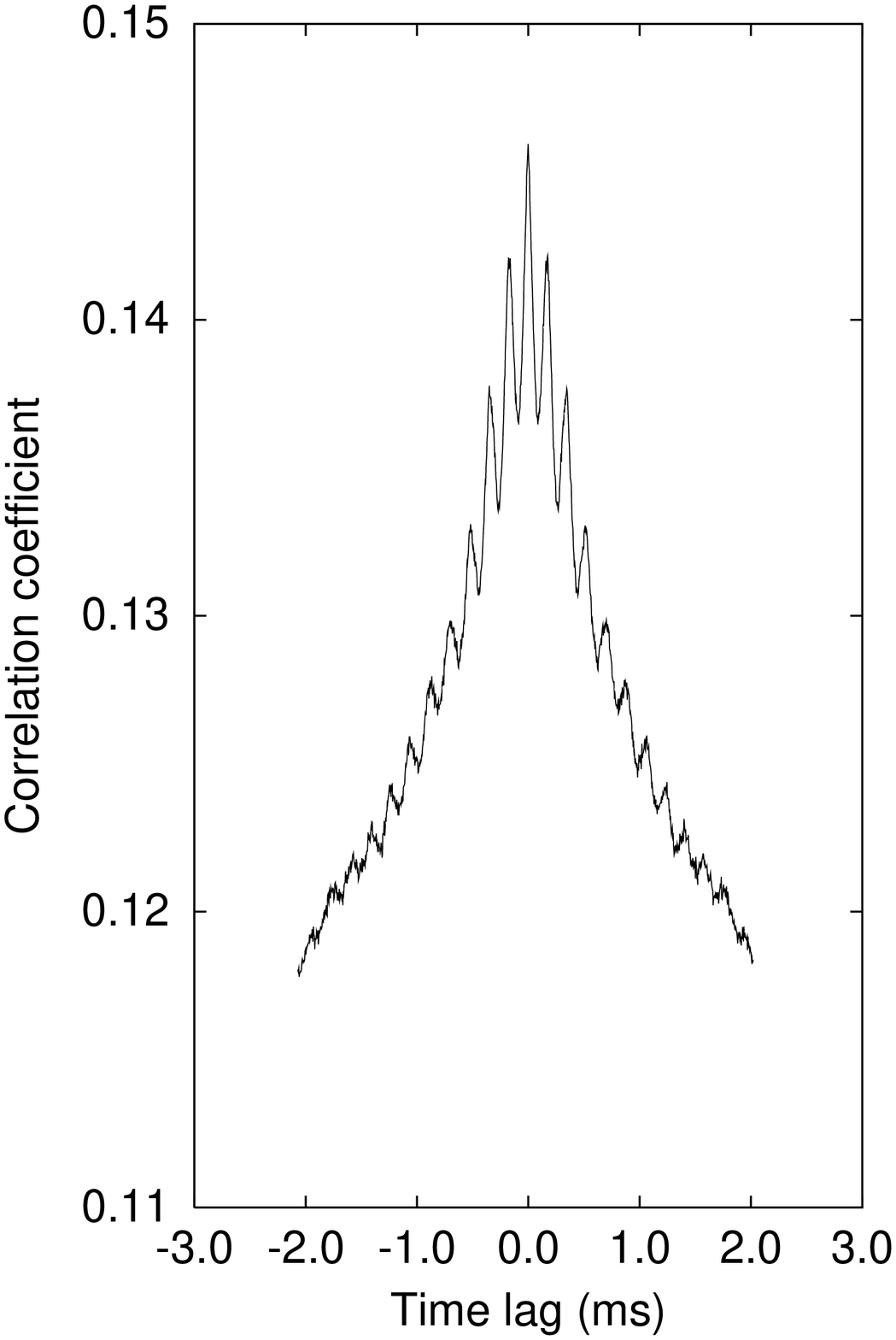}
\includegraphics[width=5cm,height=5cm]{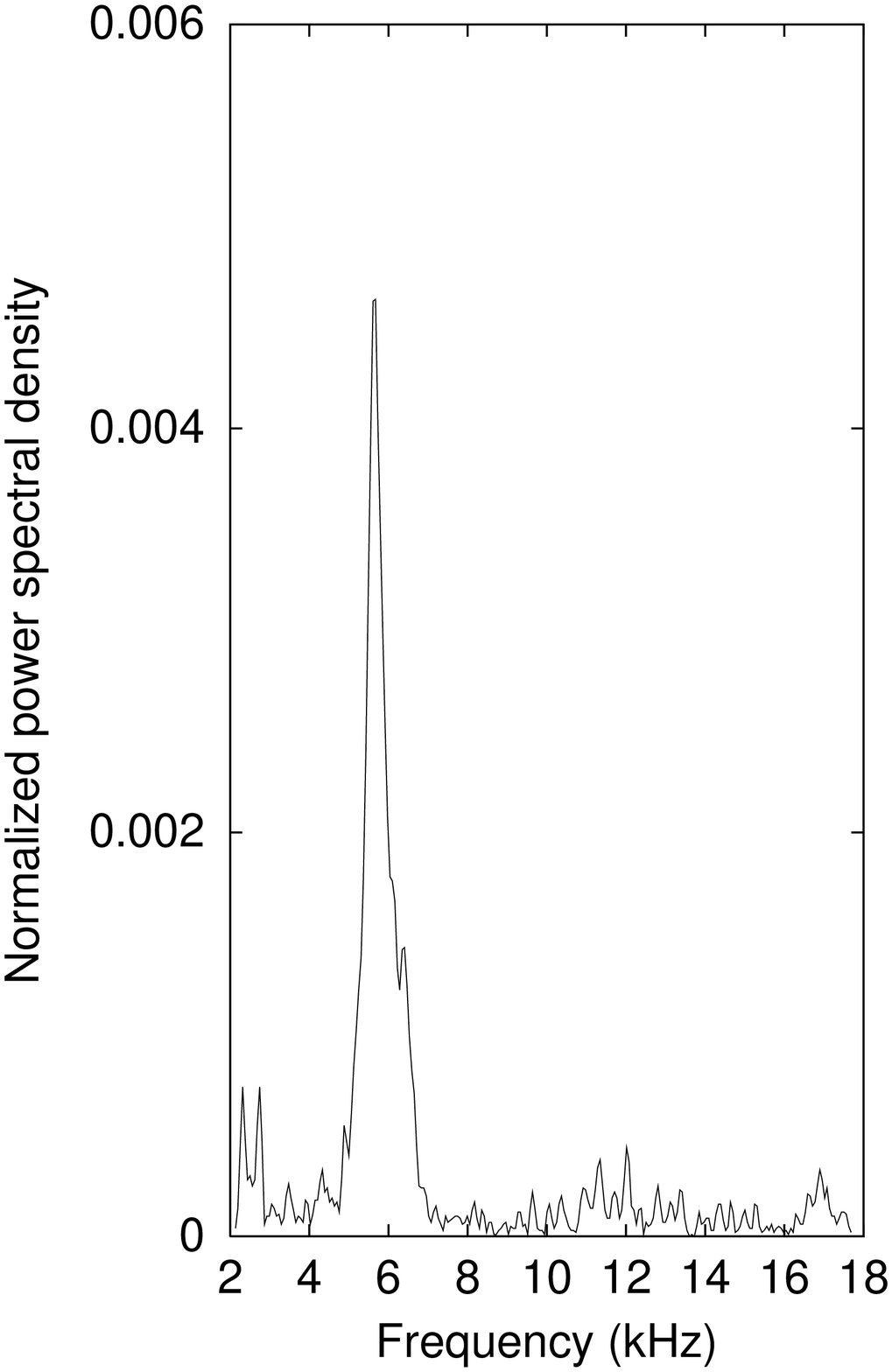}

\vskip 2mm

\caption{An example of single pulse intensities with quasi-periodic
microstructure for \object{PSR~B1133+16} (component I) recorded in the upper~(U)
and lower~(L) sidebands (left) and the corresponding CCF (middle) and cross-power
spectrum (right). For plotting purposes only, the intensities  were smoothed
with a time constant of $8~\mic$, and the CCF was smoothed with a time constant
of $1~\mic$ and displayed with an arbitrary scale. For the cross-power spectrum we plot only the frequency range above
2~kHz, since the large power spectral density from features at lower frequencies
would otherwise determine the scale completely.}

\label{example}

\end{figure*}

\section{Quasi-periodic microstructure}

\subsection{Analysis method}

\noindent
Micropulse periodicities can be revealed, e.g., by analyzing
ACFs, CCFs, or  power spectra. Since our observations were conducted in two
adjacent frequency bands, we again use CCFs and in addition cross-power spectra.

The cross-power spectrum is given as $|S_{1,2}(f_\mathrm{\mu})|^2$ with
\begin{center}
\begin{equation}
S_{1,2}(f_\mathrm{\mu})=\sum_{\tau =-T/2}^{T/2} R_{1,2}(\tau)W(\tau)\exp(-j2\pi
f_\mathrm{\mu}\tau)~,
\label{eqfifth}
\end{equation}
\end{center}
where $f_\mathrm{\mu}$ is the frequency of the microstructure periodicity,
$R_{1,2}(\tau)$ is the CCF (see equation~(\ref{eqsec})), and
$W(\tau)$ is Hanning's window~\citep{blackman1958}
used to smooth variations of $S_{1,2}(f_\mathrm{\mu})$, with $W(\tau)=\frac{1}{2}\left[1+\cos\frac{2\pi\tau}{T}\right]$.

Figure~\ref{example} shows an example of quasi-periodic microstructure within  a
single pulse of \object{PSR~B1133+16} (component I) for the upper and lower bands, the
corresponding CCF, and its cross-power spectrum.
The effect of the quasi-periodicity of the microstructure is clearly seen in the
CCF and the cross-power spectrum.

In the following analysis we will use the cross-power spectra only. That has at
least two advantages compared with the traditional use of
ACFs and CCFs. First, in case of only one strong and isolated spectral line or
feature as displayed in Figure~\ref{example} the cross-power spectrum
provides  directly numerical values for the frequency,
$f_\mathrm{\mu}$,
and width, $\Delta f_\mathrm{\mu}$,
of the detected feature that corresponds to the
quasi-periodicity. Second, in case of more than one isolated strong feature, such
features can be relatively easily distinguished in the cross-power spectrum but not so in
an ACF or a CCF. We selected a feature as ``detected'' if its power
spectral density was larger than, or equal to, 6~rms in the spectrum
on both neighboring sides of the feature, or if the power spectral density
exceeded a reasonable threshold.
The last condition was necessary to detect strong spectral features in the very
low frequency range where the local rms variation could be overestimated because of
the frequent complexity of the spectral features. In case of approximately
symmetric features, the width, $\Delta f_\mathrm{\mu}$, was
determined as the FWHM. In case of complex features we
interpreted the feature as a blend of several narrow features and estimated
the FWHM of the dominant unblended feature by
measuring the half-width at half-maximum intensity from the apparently
unblended side of the feature to the peak and then doubled that width.

In Table~\ref{psrpar1} we list the total number, $N_\mathrm{\mu-QP}$, of strong
isolated features found in the cross-power spectra of $N_\mathrm{p}$
single pulses for PSRs \object{B0950+08}, \object{B1133+16} (I,~II),
and \object{B1929+10}.

\subsubsection{Types of microstructure periodicities}

\begin{figure}[hbt]

\includegraphics[width=8cm,height=8.cm]{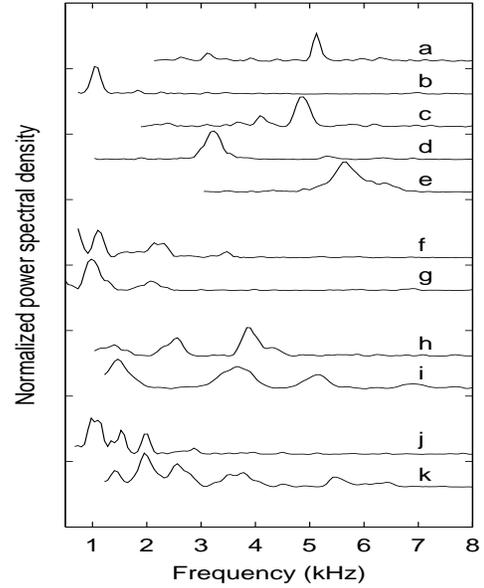}

\vskip 2mm

\caption{Four different types of cross-power spectra (see text) for selected
single pulses of \object{PSR~B1133+16} for any of its two components. Each
spectrum is normalized by the total power, i.e. by the sum of the power of all harmonics.
All plots shown in the figure start from some relatively high frequency.
Low-frequency features are much more powerful and would determine the scale of
the plots such that the high-frequency features would become almost invisible. The 
spectra were smoothed by using Hanning's window, reducing the originally high resolution only by 
$\sim 25$\%. The noise level in the plots is much smaller than the features displayed.
The spectrum (e) is a reproduction of the spectrum in the right panel of Figure~\ref{example}.}

\label{exenv}

\end{figure}

\noindent
In Figure~\ref{exenv}
we show different types of cross-power spectra for several
selected single pulses of \object{PSR~B1133+16}.
Similar types were also found for the other
two pulsars. All spectra were normalized by their total power, i.e. the sum
of all harmonics was set equal to unity.

Visual inspection shows that all spectral features
below 10~kHz are substantially broader than the applied
frequency
resolution  of about 100~Hz.
The spectra therefore differ substantially from random noise and reflect the
properties of the pulsar emission.

The spectra may be classified in four general
categories in order of increasing complexity.  The first category comprises
spectra dominated by  only one symmetrical feature (a-e).
The second category comprises spectra
with two or more isolated symmetrical features with their frequencies being
approximately multiple integers
of the lowest frequency at which an isolated feature could be identified
(f, g). These two categories correspond to the ``classic''
quasi-periodicities. However they  constitute only about 5\% of all analyzed
spectra, about the same percentage for all
three pulsars. The third category covers the spectra with several
symmetrical isolated features located at more or less random frequencies (h, i).
About 30--35\% of  all spectra fall under this category.
Finally, the fourth category comprises spectra with isolated features
at random frequencies as in the third category,
but where each isolated feature is not symmetrical but instead
structurally more complex (j, k).
This  is the most numerous category; it contains about two thirds of
all spectra of all three pulsars.

\subsection{Histograms of the microstructure period}

\begin{figure*}[t]

\hskip 15mm\includegraphics[width=6cm,height=6cm]{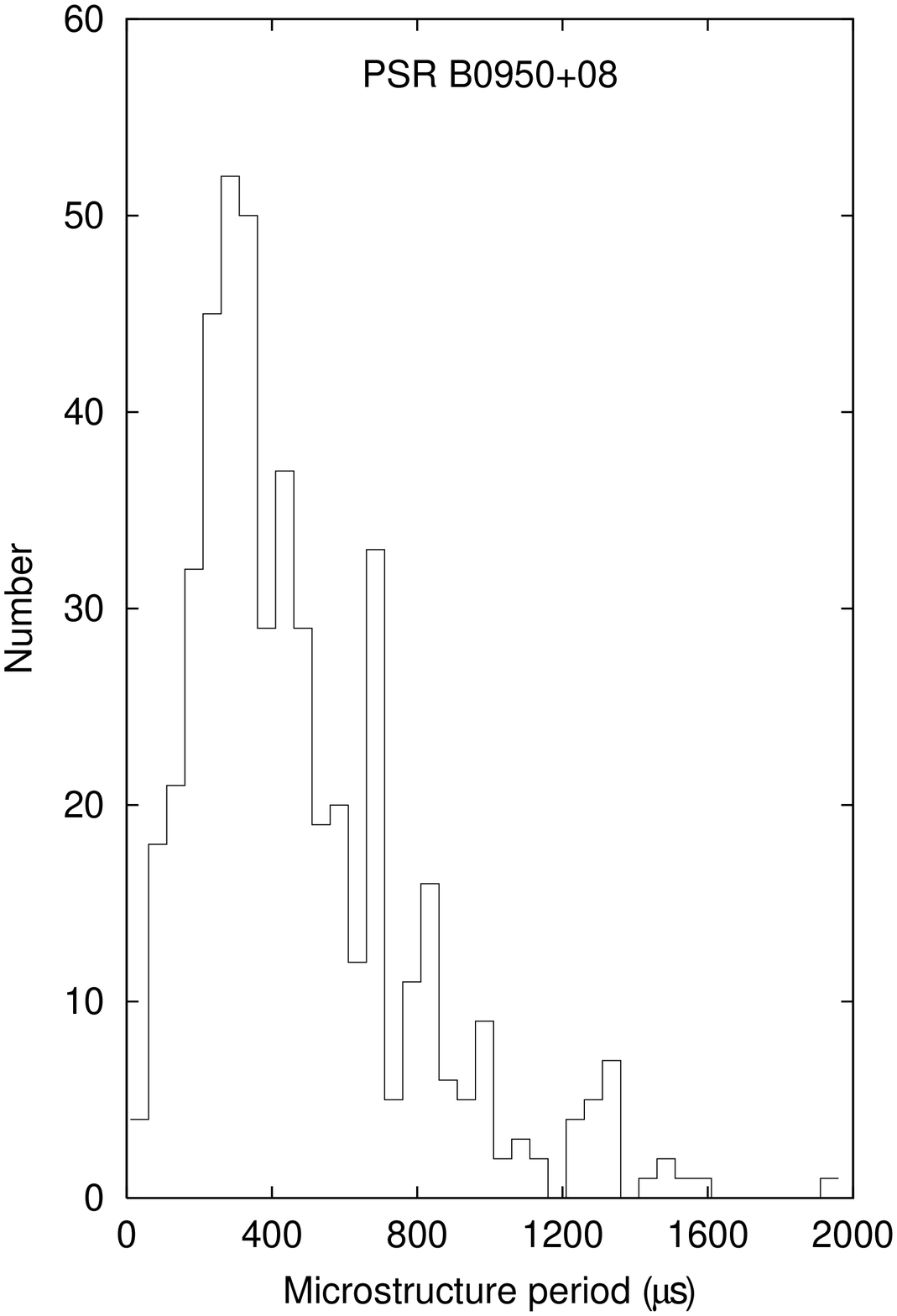}
\hskip 10mm\includegraphics[width=6cm,height=6cm]{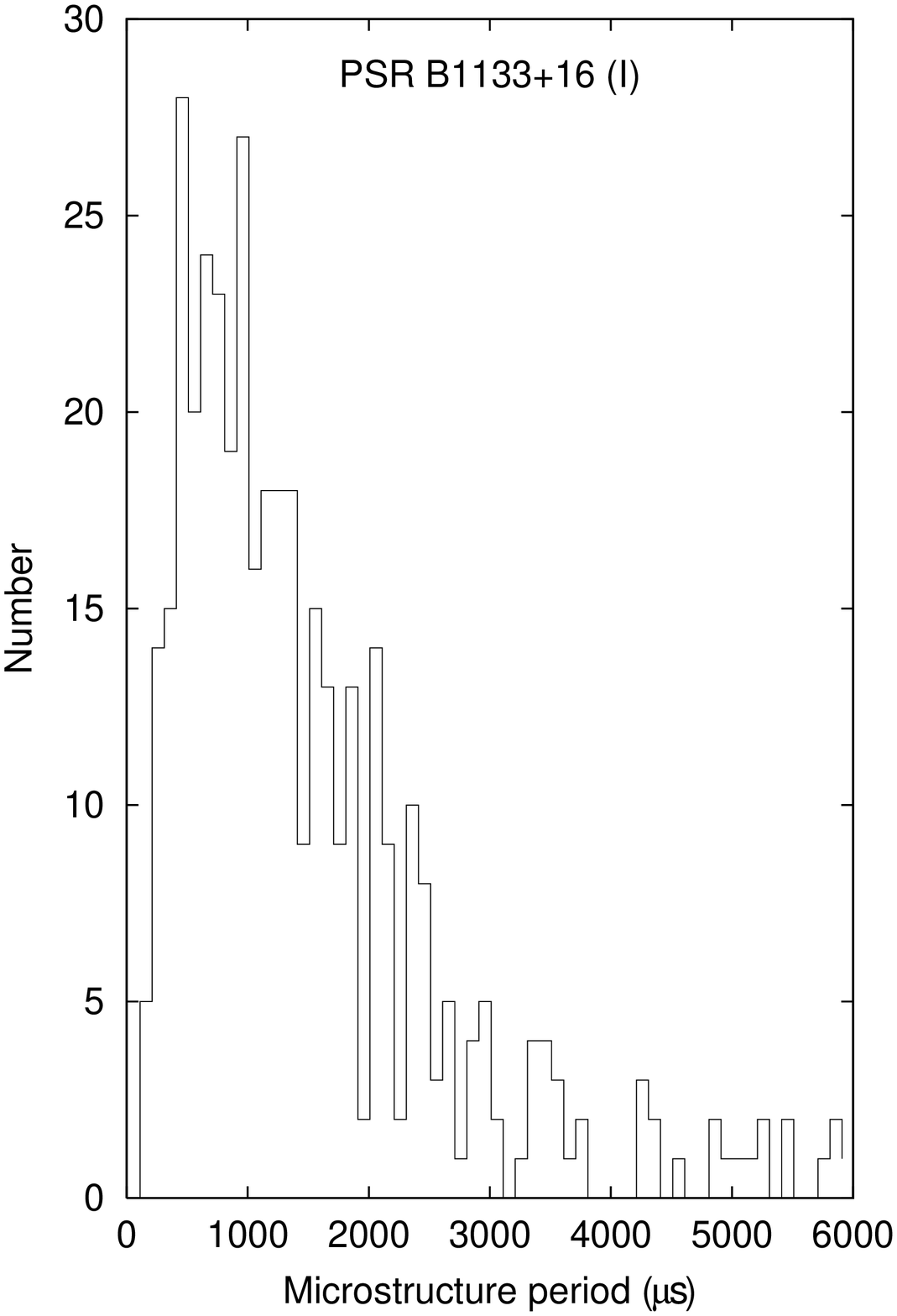}

\hskip 15mm\includegraphics[width=6cm,height=6cm]{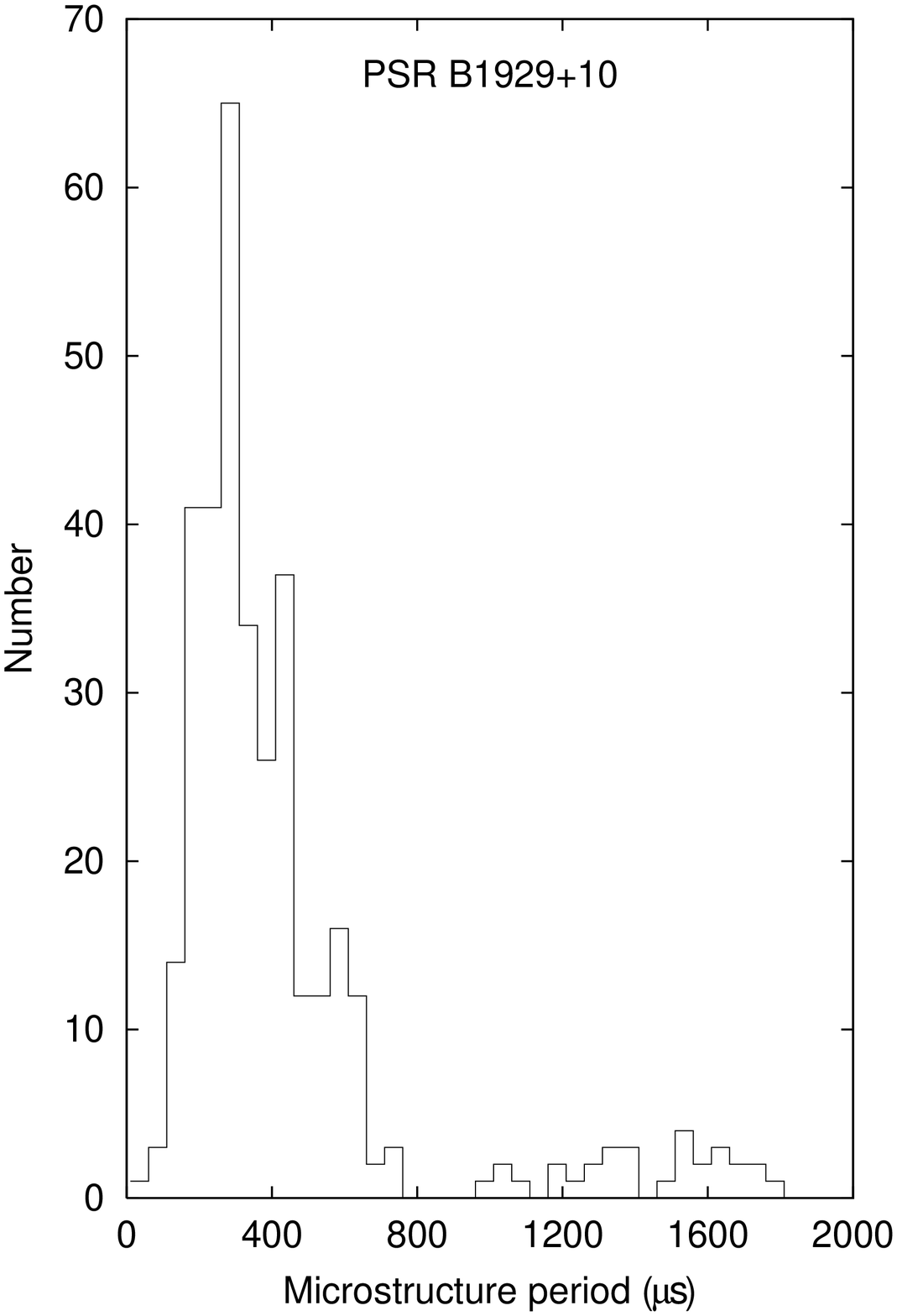}
\hskip 10mm\includegraphics[width=6cm,height=6cm]{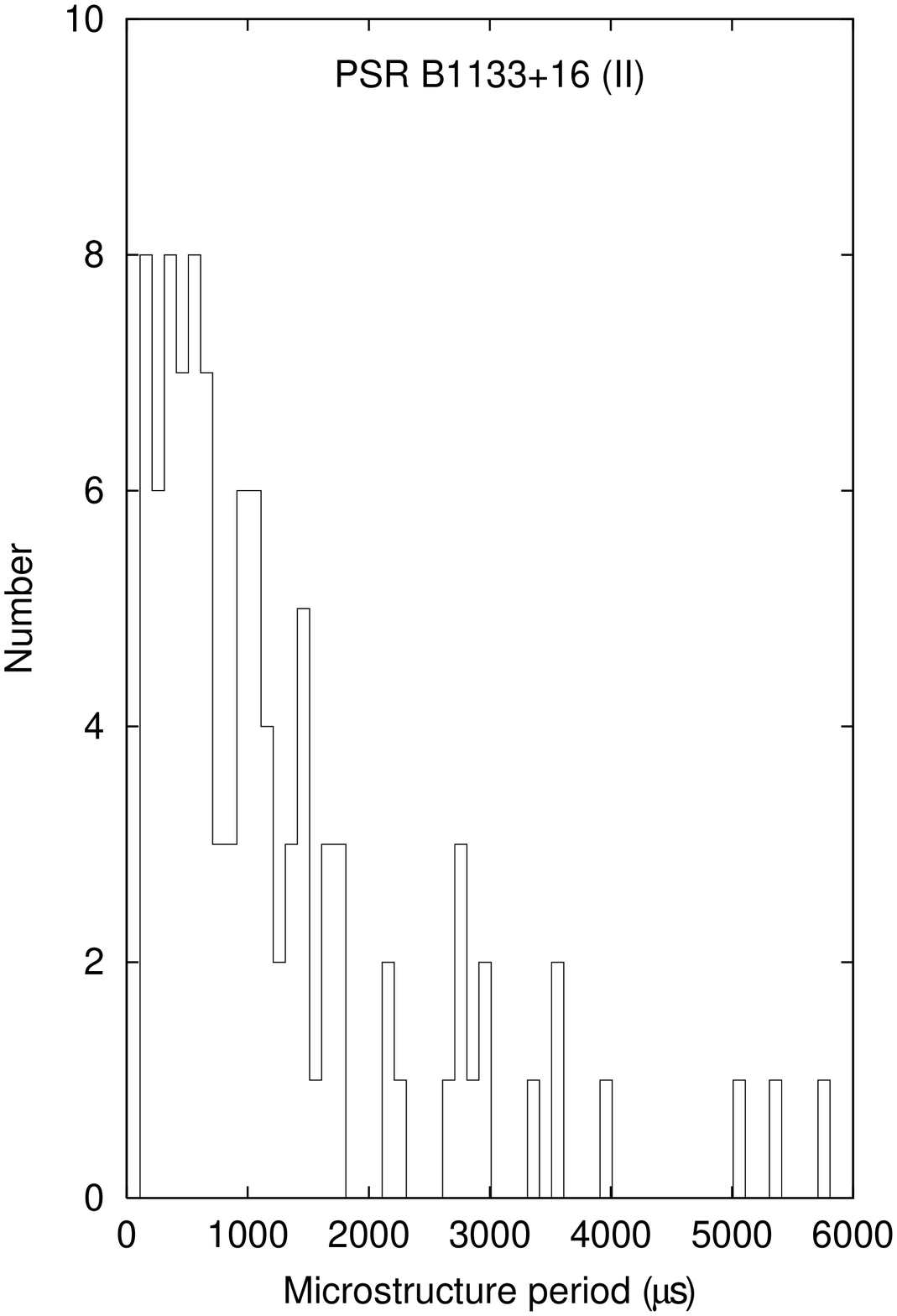}

\vskip 2mm

\caption{Histograms of the microstructure periods, $P_\mathrm{\mu}$, for the
quasi-periodic features detected in the cross-power spectra.}

\label{periods}

\end{figure*}

\noindent
In  Figure~\ref{periods} we show
the histograms of the microstructure period,
$P_\mathrm{\mu}$  ($P_\mathrm{\mu}=1/f_\mathrm{\mu}$),
for PSRs \object{B0950+08}, \object{B1133+16} (I,~II)
and \object{B1929+10}. There are some general similarities between them.
The histograms are skewed towards smaller microstructure periods.
They have a relatively narrow peak at about $300~\mic$ for PSRs
\object{B0950+08} and \object{B1929+10} and a broader peak at
$\sim 800~\mic$ for component I of \object{PSR~B1133+16} and  $\sim 400~\mic$
for component II, two to fourfold larger than $\tau_\mathrm{\mu-broad}$. The histograms
fall off sharply towards smaller periods.
The widths of the histograms are relatively large. The bulk of microstructure
periods (75\%) falls in the range of 0.1--1.0~ms for \object{PSR~B0950+08},
0.2--3.0~ms for component I of \object{PSR~B1133+16}, 0.1--2.0~ms
for component II of \object{PSR~B1133+16}, and 0.2--0.7~ms for
\object{PSR~B1929+10}.
For PSRs \object{B0950+08} and \object{B1133+16} for which microstructure
results were published earlier,
these range values agree well with those obtained by others at other frequencies
with a smaller sample size~\citep{hankins1971, soglasnov1981,
soglasnov1983, cordes1990, smirnova1994}.

Again, for \object{PSR~B1133+16} there is a clear difference in the histogram of
microstructure periods.
\begin{figure*}[t]

\hskip 15mm\includegraphics[width=6cm,height=6cm]{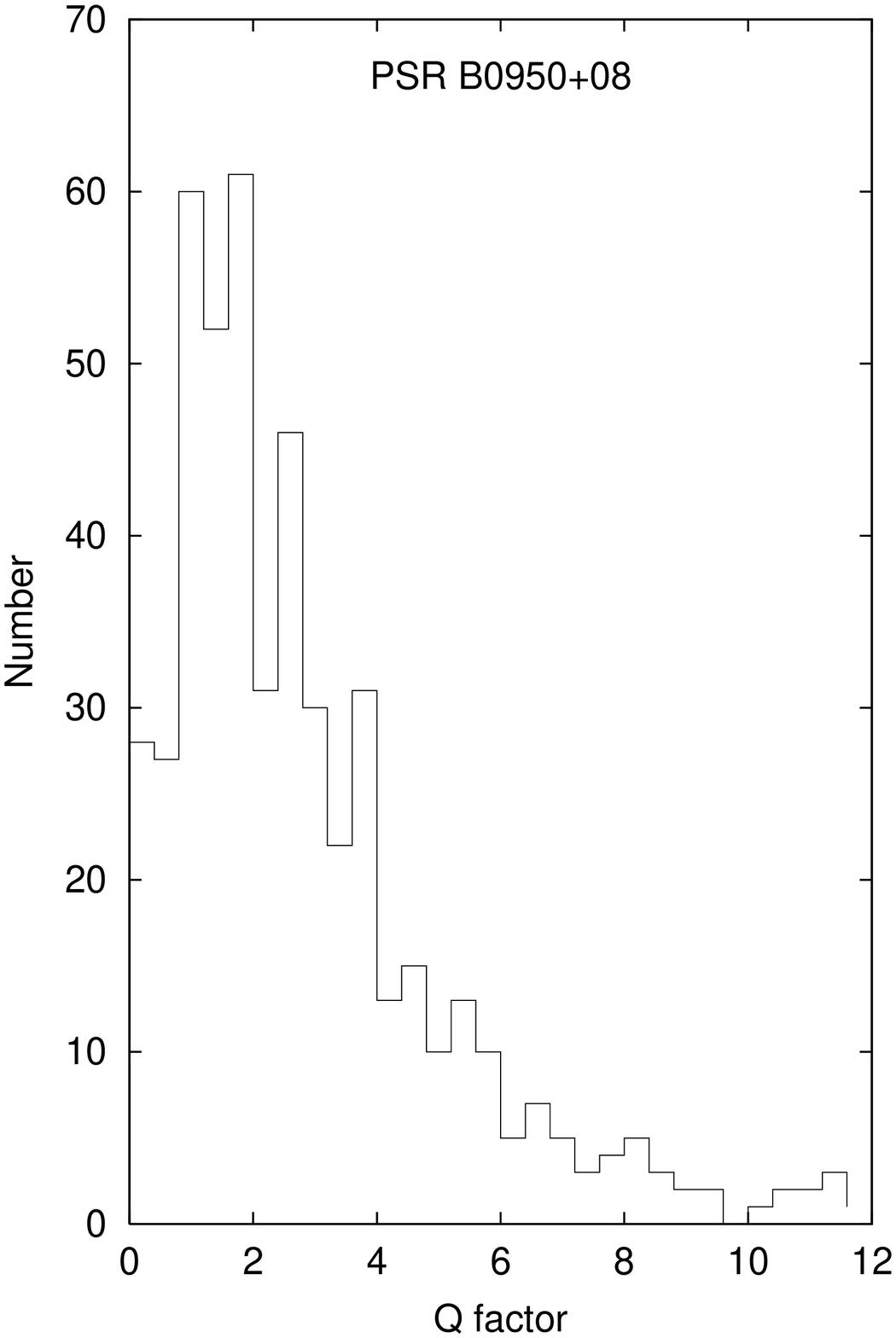}
\hskip 10mm\includegraphics[width=6cm,height=6cm]{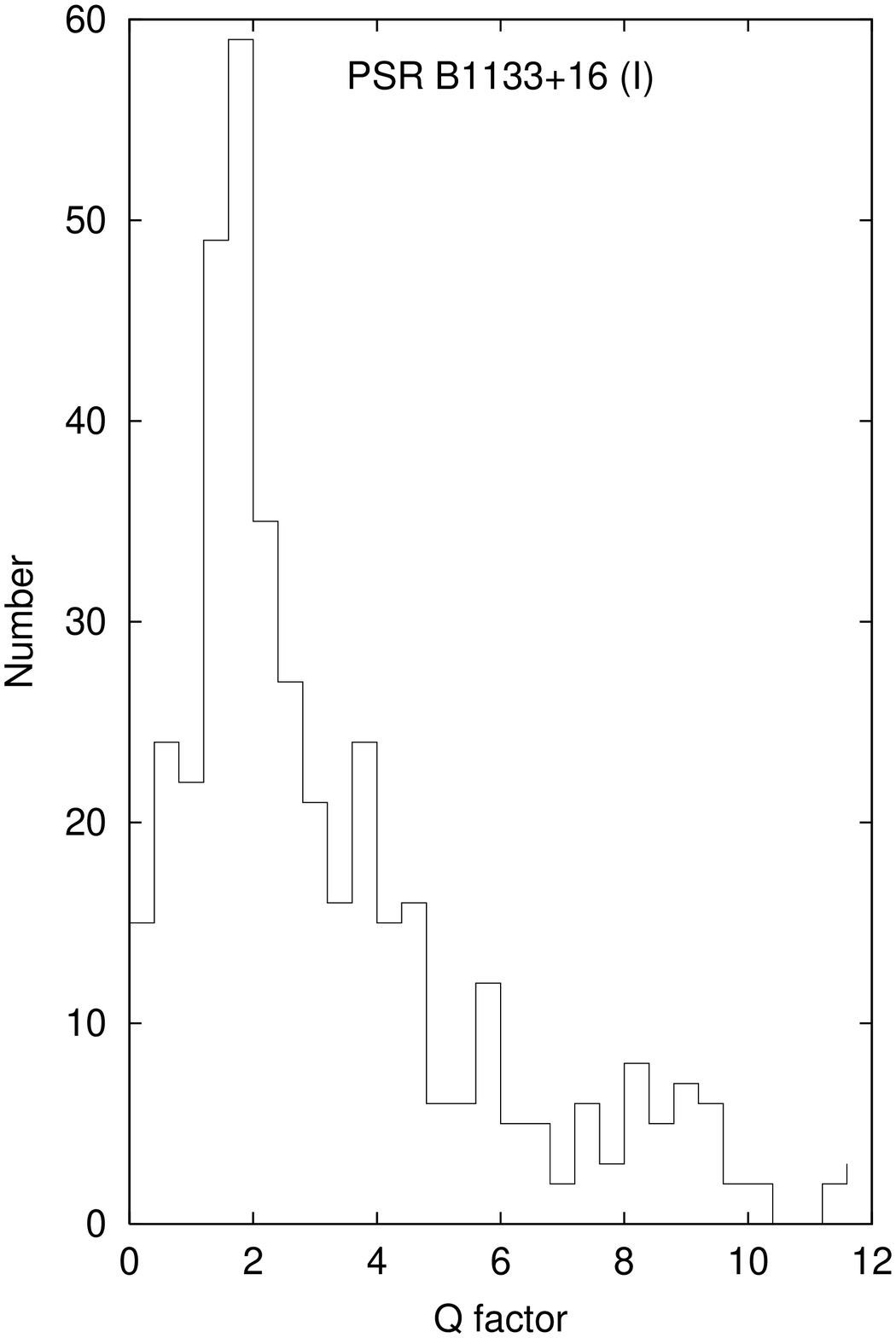}

\hskip 15mm\includegraphics[width=6cm,height=6cm]{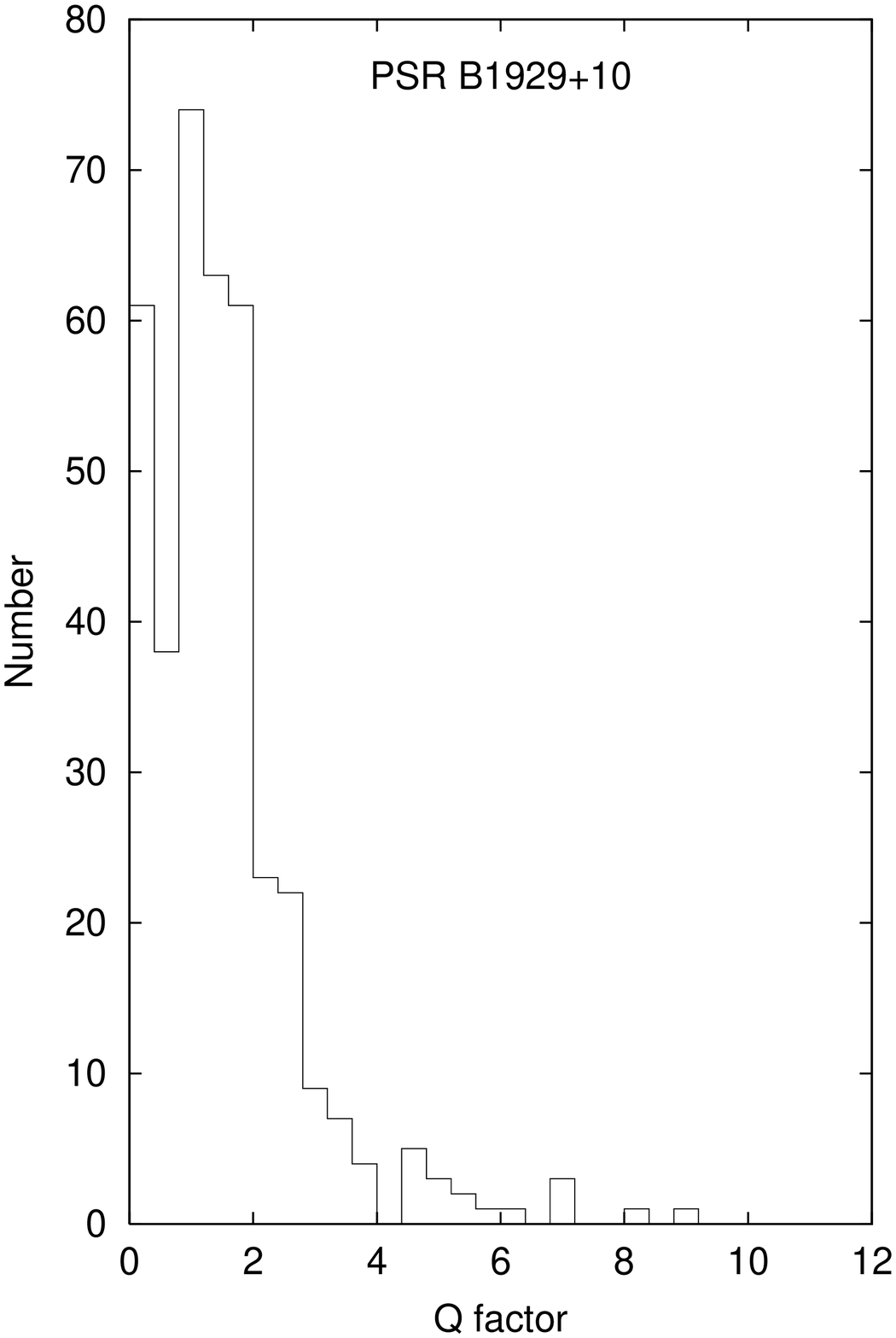}
\hskip 10mm\includegraphics[width=6cm,height=6cm]{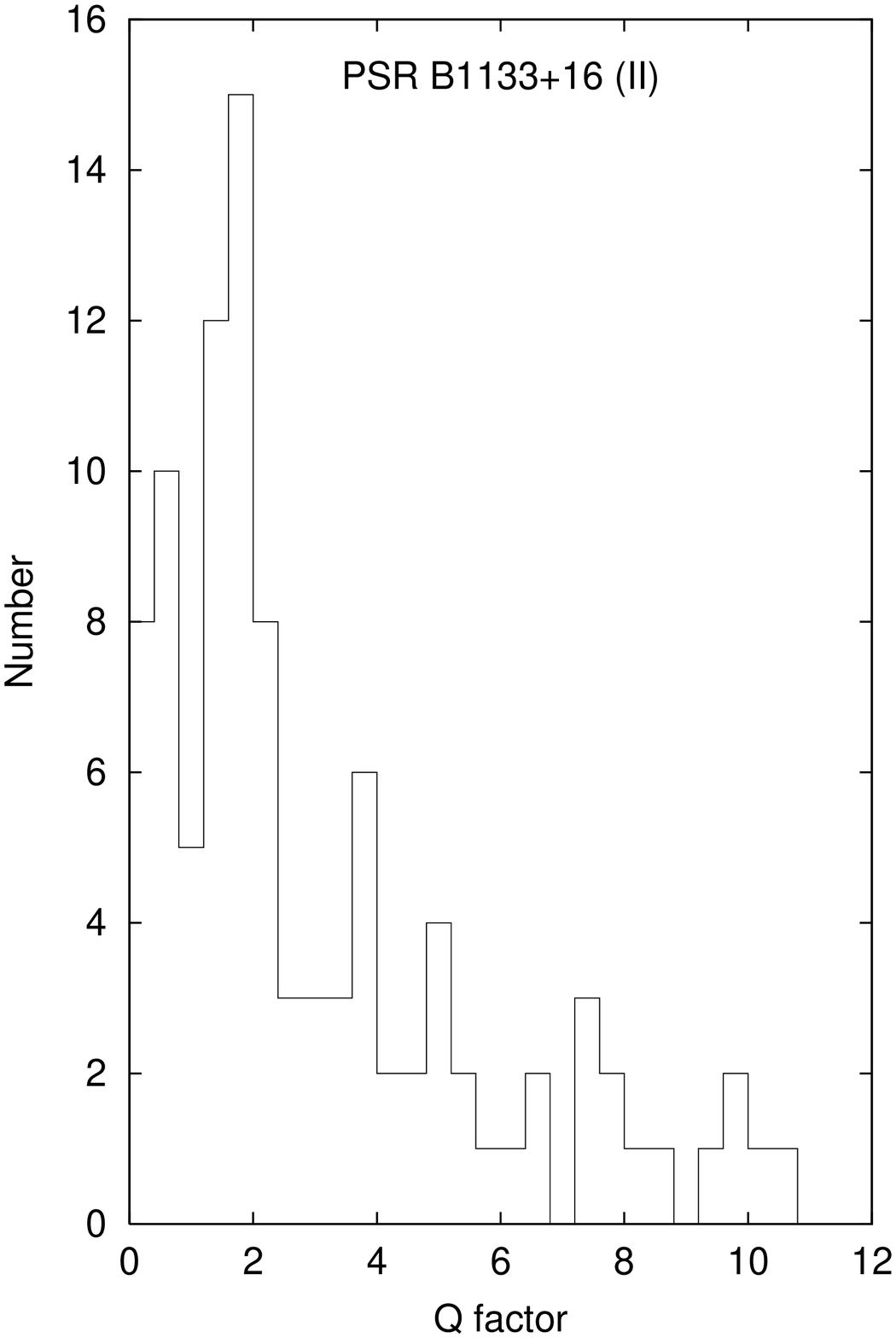}

\vskip 2mm

\caption{Histograms of the microstructure period $Q$-factor.}

\label{qfactor}

\end{figure*}
Quasi-periodicities with the shortest periods  are relatively more numerous
for the second component than for the first component.

\subsection{Histograms of the $Q$-factor of the microstructure periods}
\label{Q-factor}

\noindent
The microstructure periods were determined from
the peaks of the individual
features in the cross-power spectra at frequency $f_\mathrm{\mu}$.
The widths, $\Delta f_\mathrm{\mu}$, of the features varied widely.
To determine the sharpness of the features or the ``fuzziness''
of the microstructure periods we followed~\citet{cordes1990}
and defined the $Q$-factor as $Q=f_\mathrm{\mu}/\Delta f_\mathrm{\mu}$.
Figure~ \ref{qfactor} shows the histograms of the $Q$-factor  for each of the
three pulsars. A large majority of  spectral features have relatively
low values of $Q$ with the most frequent value of about 1.5 to 2.
In 75\% of all cases $Q<5$ for \object{PSR~B0950+08} and each of the components of
\object{PSR~B1133+16}, and $Q<4$  for \object{PSR~B1929+10}.
However, some relatively high values with $Q\sim 10$ were also found,
but only for \object{PSR~B0950+08} and the two components of \object{PSR~B1133+16}.

\subsection{Average cross-power spectra}

\begin{figure}[hbt]

\includegraphics[width=8.5cm, height=9cm]{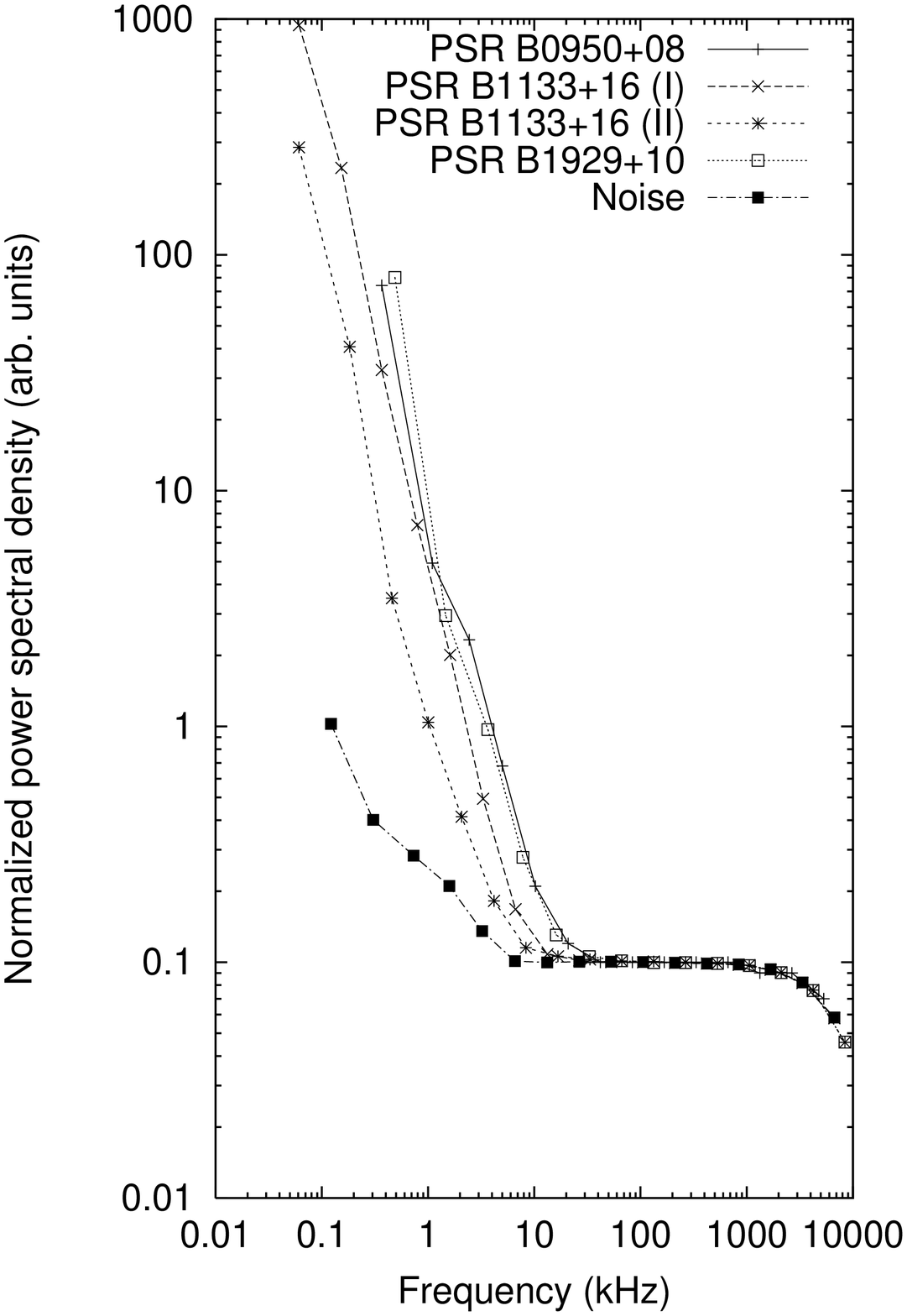}

\vskip 2mm

\caption{Average cross-power spectra of the intensities at two conjugate 16-MHz
bands for ON-pulse and OFF-pulse emission. The spectra were normalized to the
largely flat part of the OFF-pulse spectrum between 50~kHz and 1\,000~kHz.}

\label{avlogspectr}

\end{figure}

\noindent
In order to investigate the general aspects of the cross-power spectra we
averaged the individual cross-power spectra
for each of PSRs \object{B0950+08} and \object{B1929+10} and each component
of \object{PSR~B1133+16}.
We then normalized each of the spectra so that they have equal
power spectral densities at  frequencies higher than about 50~kHz
where they are largely flat and identical to the
equivalent spectra obtained for the OFF-pulse signal. In
Figure~\ref{avlogspectr} we show these
spectra for all three pulsars and compare them with the OFF-pulse spectra.
At frequencies below about 20~kHz the ON-pulse power spectra
show a large excess above the noise power spectrum. This excess reflects the
presence of microstructure fluctuation power with timescales
down to $\sim 50~\mic$. The value corresponds to the
timescales of microstructure revealed through the analysis of individual CCFs.
The low frequency part can be approximated  by the
power law, $S\propto f^{-\alpha}$, with the exponent $\alpha \sim2$
for all three  pulsars. In comparison, the exponent for the OFF-pulse spectrum
is about 0.5.
Such an increase of power toward the low frequencies is typical for 
low-noise amplifiers. The noise with these characteristics is called excess noise, 
low-frequency noise, or $1/f$ noise. Usually its exponent
is in the range from 0.8 to 1.5. Our rather low value of the exponent
is probably due to the properties of the cross-power spectrum in our analysis.
 Clearly, the analysis  of the average power spectrum is another
way to detect the presence of microstructure and its quasi-periodicities.

\section{Discussion}
\label{discuss}

We have presented a statistical analysis of the properties of microstructure
for \object{PSR~B0950+08}, the two components of \object{PSR~B1133+16},
and \object{PSR~B1929+10}, the most extensive such analysis with one of
the highest time resolutions  (62.5~ns) ever made.

Microstructure with a characteristic broad timescale between $\sim 95$ and
$\sim 450~\mic$ in the ACF or CCF
was confirmed for PSRs \object{B0950+08} and  \object{B1133+16}  and for the
first time measured for \object{PSR~B1929+10}. Microstructure with a characteristic
short timescale of only $\sim 10~\mic$ was found for \object{PSR~B0950+08}, the second
component of \object{PSR~B1133+16}, and for \object{PSR~B1929+10}, the first
time such a short characteristic timescale was detected in the average
ACF or CCF from any pulsar.
Clearly, microstructure can have broad as well as short timescales for the same
pulsar. This conclusion is further supported by the histograms of the microstructure timescales
which generally showed a moderately
rising part toward shorter timescales, corresponding to the
broad micropulses, and a sharply rising part corresponding to short micropulses.

Different characteristic timescales for micropulses were previously found
by others, albeit for much broader widths. \citet{soglasnov1981} observed PSRs
\object{B0809+74} and \object{B1133+16} at 102.5~MHz with a time resolution of $10~\mic$
and found  that a long timescale of $\geq 500~\mic$ and a shorter one of
$\leq 500~\mic$ are present in the ACFs of these pulsars. Later it was shown for
the same pulsars  that micropulses
with a long timescale are clearly correlated over a broad frequency range of
70--102.5~MHz~\citep{popov1987} while short timescale micropulses are not
correlated over a narrow frequency range of 1.6~MHz at a center frequency of
102.5~MHz~\citep{smirnova1986}.
 In contrast, in the present study we have found for PSRs \object{B1133+16}
 (component II) and \object{B1929+10} a clear correlation between the short timescale
microstructure in the two adjacent bands separated by 16~MHz or 1\% of the center
frequency.

Despite several hours of observations, no nanopulses were detected for any of
the pulsars. The shortest micropulse was found for the second component of
\object{PSR~B1133+16} and had a width of $2~\mic$.
In particular, the histograms of the microstructure widths for
all three pulsars showed a sharp cutoff at 5 to 10~$\mic$. The rare occurrence
of micropulses with short widths of only a few microseconds
but with flux densities strong enough to be detected is consistent
with earlier results for \object{PSR~B1133+16} at nearly the same frequency
by~\citet{bartel1978} and~\citet{bartel1982}  who also found only one or at best
a few micropulses with such a short width after several hours of observations.
Either micropulses with widths shorter than about $1~\mic$ do not exist at our
observing frequency in the three pulsars we studied or they are so weak or
occur so infrequent that they have not yet been observed.

In contrast to our pulsars,~\citet{hankins2000}
reported to have found extremely strong fluctuations for the Crab pulsar
which are still unresolved with a time resolution of 10~ns. It remains to be
seen whether the difference in the shortest timescale for
microstructure for our three pulsars on the one hand
and for the Crab pulsar on the other hand is caused by the Crab pulsar's young
age, small period, or general uniqueness.
Observations of other pulsars with a similarly high time
resolution as ours are needed to investigate further
this aspect of microstructure research.

Detailed inspection of the central portion of the average CCFs showed that the
time delay of the micropulses between the two sidebands is for
\object{PSR~B1929+10} within the error as expected,
but is for PSRs \object{B0950+08} and \object{B1133+16} 1 to 2\% shorter than expected.
The latter discrepancy has never been observed before.
Earlier investigations of a possible discrepancy  were based on observations
with a coarser timing accuracy and a larger relative and absolute frequency
difference of the two channels than ours~\citep{boriakoff1983, popov1987, sallmen1999}.

The quasi-periodicities observed for all three pulsars were in most cases only
relatively weakly pronounced with $Q$-factors generally between 1.5
and 2 and cross-power spectra showing rather complex features.
Only in 5\% of all cases did we detect
classic quasi-periodicities with single spectral features and their harmonics.
In fact, it appears that the complexity of the spectra almost continuously
covers all of our four different categories. In this sense there
is no  fundamental difference between
spectra of classic quasi-periodicities and spectra with no clear periodicities
at all.

In this context the interpretation of microstructure quasi-periodicities
in terms of  modes of vibrations of neutron stars~\citep{boriakoff1976,
vanhorn1980, hansen1980}
is not attractive although some modes fall well into the range of the  observed
quasi-periodicities~\citep[see][]{mcdermott1988}.
However, our values of the  $Q$-factor
are very small in comparison to about 1\,000 expected from stellar
vibrations. Further, the essential continuum of the degree of complexity
of the cross-power spectra and the lack of any resonance spike in the
histograms for the microstructure period, $P_\mathrm{\mu}$,
make it very unlikely that the
microstructure quasi-periodicities are caused by vibrations of
the neutron star. Similar conclusions were drawn by~\citet{cordes1990}
as a result of a study of five pulsars at several frequencies.

We find it most remarkable that the micropulses in the two components of
\object{PSR~B1133+16} display different characteristics in every aspect we
analyzed. The average CCFs, the histograms of microstructure timescales,
the histograms of microstructure period, and
the average cross-power spectra are all significantly different for the
two components. Most striking is the more dominant occurrence of short
micropulses in the second component which is reflected in the sharp spike
of the central portion of its average CCF and the twofold higher
frequency of occurrence in the histogram for $\leq 100~\mic$ widths.

Differences in microstructure parameters for the two components of
\object{PSR~B1133+16} were first
reported by~\citet{cordes1977} in their study of polarization properties
of microstructure at 430 MHz with a time resolution of $8~\mic$. Also,
~\citet{smirnova1994} found that in the range from 200 to
$1\,500~\mic$ micropulses of component II more frequently had smaller
widths than those of component I in their observations at 1.4~GHz with a
time resolution of $150~\mic$. These results are at least qualitatively
consistent with our result.

Are the micropulses  we observe due to a longitudinal emission modulation and
therefore to a sweep of the beam past the observer, or due to a radial
emission modulation and therefore to the intrinsic time  variability of the
emission, or perhaps due to a combination of both? The average pulse profile
clearly reflects the beam pattern; the equivalent profile width is
proportional to P~\citep{lyne1988}.  Subpulses in single pulses have a typical
width $\propto \mathrm{P}^{0.8}$ indicating that the millisecond
structure like the average pulse profile can largely be
interpreted as being due to the sweep of the beam past the
observer~\citep{bartel1980}.
Broad micropulses of several $100~\mic$ duration were reported to
have a somewhat weaker dependence on period ($\propto \mathrm{P}^{0.5}$;~\citealt{taylor1975};
see also~\citealt{ferguson1977}) reflecting perhaps already part of the
elementary emission mechanism with its temporal variations.

Three pulsars  are not enough to test this relation
since the scatter about the regression line is relatively large. However,
we can compare our broad, narrow, and shortest microstructure widths  with the
component widths of the average profile, $t_{1/2}$, and with the subpulse
widths, $\tau_\mathrm{m}$, listed for the three pulsars in Table~\ref{psrpar2}.
The ratios for the comparison are given in Table~\ref{psrpar3}. While the
microstructure widths relative to P can indeed vary within their categories by
several times their errors , they are constant within about twice the combined errors relative to
$t_{1/2}$ and $\tau_\mathrm{m}$ (with the exception of $\tau_\mathrm{\mu-broad}$ relative to $t_{1/2}$ for
component I of PSR B1133+16).

\begin{table*}

\caption{Comparison of pulsar parameters and $\gamma$ factors.
The parameters for comparison in columns 2--4 are taken from Table~\ref{psrpar2}. For
PSR B0950+08 3\% errors are assumed for $t_{1/2}$ and $\tau_m$. The
parameter $\gamma_\mathrm{\mu-narrow}^{(\zeta=90^o)}$ is derived for an angle,
$\zeta$, between the line of sight and the rotation axis of the pulsar of $90^o$
and $\gamma_\mathrm{\mu-narrow}$ for an angle estimated from the polarization
angle sweep across the beam for the hollow-cone model, $\zeta=10.1$, $55.0$, $10.0$
for PSRs B0950+08, B1133+16 (II), and B1929+10, respectively~\citep{lyne1988}.
For the first two pulsars the values for $\zeta$ are similar
to those given by~\citet{rankin1993}. For PSR B1929+10~\citet{rankin1993} interpreted the viewing
geometry very differently in comparison to~\citet{lyne1988}, but did not give
a value of $\zeta$. The errors in columns 5--6 are ``symmetrized'' errors based on the errors of
$\tau_\mathrm{\mu-narrow}$ only.}

\label{psrpar3}

\begin{flushleft}
\begin{tabular}{lcccccc}
\hline
{\Large\strut}PSR B                                   & ${\tau_\mathrm{\mu-broad}}$                           &
${\tau_\mathrm{\mu-narrow}}$                          & ${\tau_\mathrm{\mu-shortest}}$                        &
$\gamma_\mathrm{\mu-narrow}^{(\zeta=90^o)}$           & $\gamma_\mathrm{\mu-narrow}$                          &
$c\cdot\tau_\mathrm{\mu-narrow}$ \\
{\Large\strut}                                        & $(10^{-4}~\mathrm{P}\phantom{1\!2})$                  &
$(10^{-5}~\mathrm{P}\phantom{1\!2})$                  & $(10^{-3}~\tau_\mathrm{m})$                           &
                                                      &                                                       &
(km)       \\
{\Large\strut}                                        & $(10^{-2}~t_{1/2})$                                   &
$(10^{-3}~t_{1/2})$                                   &                                                       &
                                                      &                                                       & \\
{\Large\strut}                                        & $(10^{-2}~\tau_\mathrm{m}\phantom{1})$                &
$(10^{-3}~\tau_\mathrm{m}\phantom{1})$                &                                                       &
                                                      &                                                       & \\
\hline
{\Large\strut}0950+08                                 & $5.3 \pm 0.2$                                         &
$ 5.5 \pm 1.2 $                                       & $1.40 \pm 0.04$                                       &
$\phantom{1}2\,900\pm 700\phantom{0}$                 & $16\,400 \pm 4\,000$                                  &
$4.2\pm 0.9$               \\
                                                      & $1.5 \pm 0.1$                                       &
$1.5 \pm 0.3$                                         &                                                       &
                                                      &                                                       &  \\
                                                      & $2.7 \pm 0.1$                                         &
$2.8\pm 0.6$                                          &                                                       &
                                                      &                                                       &  \\
{\Large\strut}1133+16 (I)                             & $3.6 \pm 0.3$                                         &
$\phantom{2}-\phantom{2}$                             & $\phantom{2}-\phantom{2}$                             &
$\phantom{2}-\phantom{2}$                             & $\phantom{2}-\phantom{2}$                             &
$\phantom{2}-\phantom{2}$  \\
                                                      & $7.2 \pm 0.9$                                         &
$\phantom{2}-\phantom{2}$                             &                                                       &
                                                      &                                                       &  \\
                                                      & $\phantom{2}-\phantom{2}$                             &
$\phantom{2}-\phantom{2}$                             &                                                       &
                                                      &                                                       &  \\
{\Large\strut}1133+16 (II)                            & $0.9 \pm 0.2$                                         &
$ 0.9 \pm 0.3 $                                       & $\phantom{2}-\phantom{2}$                             &
$16\,200\pm 5\,000$                                   & $19\,700\pm6\,000$                                    &
$3.3\pm 0.9$               \\
                                                      & $1.0 \pm 0.2$                                         &
$1.0 \pm 0.3$                                         &                                                       &
                                                      &                                                       &  \\
                                                      & $\phantom{2}-\phantom{2}$                             &
$\phantom{2}-\phantom{2}$                             &                                                       &
                                                      &                                                       &  \\
{\Large\strut}1929+10                                 & $4.2 \pm 0.4$                                         &
$ 4.0 \pm 1.3 $                                       & $1.47 \pm 0.04$                                       &
$\phantom{1}4\,000\pm1\,500$                          & $23\,000\pm9\,000$                                    &
$2.7 \pm 0.9$              \\
                                                      & $1.6 \pm 0.2$                                         &
$1.5 \pm 0.5$                                         &                                                       &
                                                      &                                                       &  \\
                                                      & $2.8 \pm 0.3$                                         &
$2.7\pm0.9$                                           &                                                       &
                                                      &                                                       &  \\
\hline

\end{tabular}
\end{flushleft}

\end{table*}

In other words, with the exception of the broad micropulse widths for
\object{PSR~B1133+16} (I),
each of the different types of micropulse width scales within
$\sim2\sigma$ with the subpulse width and the average pulse component width. That may be
an intriguing result. It therefore appears possible that
even the fastest observable intensity fluctuations
still have a strong component  of the longitudinal emission modulation and are
at least partly due to the sweep of the beam.
If so, we can compute the factor $\gamma$ as the energy relative to the rest
energy of the particles moving along the curved magnetic field lines and beaming
in the direction of motion. For point sources,
\begin{equation}
\gamma = \frac{1}{\phi\cdot\sin\zeta} = \frac{\mathrm{P}}{2\pi\sin\zeta\cdot\tau_\mathrm{\mu-narrow}}
\end{equation}
\citep[see][]{lange1998}.
Here, $\phi$ is $\tau_\mathrm{\mu-narrow}$ in radians relative to P, and $\zeta$ is
the angle between the rotation axis of the pulsar and the line of sight to the
observer. For the conservative assumption of $\zeta=90^o$ the values of $\gamma$
differ by up to a factor of 5. However, if the angle $\zeta$ is taken from
estimates obtained from the sweep of the polarization angle across the pulse
profile~\citep{lyne1988}, the values for $\gamma$ are much more
uniform with a mean of about 20\,000 and a spread of not more than a factor 1.4 which is well within
the errors.
Perhaps such a high value for $\gamma$ is representative for the relativistic
energy of the particles.

Any longitudinal emission pattern of the beam on this small angular scale
is likely related to the radial pattern through the radius to
frequency relation and the curved open magnetic field lines of the polar
regions. For instance, for \object{PSR~B0950+08} the subpulse ACF width increases
from 4.2~ms at 2.7~GHz to 5.0~ms at 1.7~GHz~\citep{bartel1980}, reflecting the above
frequency dependent geometry. Since the subpulses
can be approximated to first order as Gaussians\footnote{A detailed analysis
showed that the average ACF of subpulses of a representative sample of pulsars,
and by implication the average subpulse itself, is sharper at the top than
either a Gaussian or a Lorenzian but flatter than a Gaussian and similar to a
Lorenzian in the wings~\citep{bartel1980}.}, the 50\% ACF width has to be scaled upwards by
a factor of 1.4 to yield the FWHM of the subpulses. If we had observed the
subpulses over a continuous
bandwidth from 2.7 to 1.7~GHz, they would be smeared in time at the wings by
about a third of the  difference
of the scaled widths or 0.37~ms. On average, the smearing time for a subpulse
would be about 0.19~ms. Correspondingly, we could expect our micropulses
observed over a total of the two bands of 32~MHz
to be smeared in time on average by about $6.1~\mic$.
For PSRs \object{B1133+16} and \object{B1929+10} the expected smearing times
based on the differences of the scaled ACF widths are 4.5 (see caption of
Table~\ref{psrpar2} for \object{PSR~B1133+16}) and $3.7~\mic$, respectively.
The expected smearing times for the first and third pulsar are in good
agreement with their shortest widths. For \object{PSR~B1133+16} the ACF width
refers to a weighted mean of the ACF widths of both components.
The mean of our shortest widths for both components is $4.5~\mic$ which
equally well agrees with the expected smearing time for both components.

In this context the residual delay of $\sim2~\mic$ that we observed for the
microstructure of the two bands for two pulsars
can be interpreted as being due to the curved geometry of the magnetic field
lines and to the radius to frequency mapping. One might expect the delay to be a
function of the phase of the micropulses in the pulse window. Micropulses in the
leading part of the pulse window would first appear at the lower frequency,
micropulses in the trailing part first at the
higher frequency. In case of \object{PSR~B1133+16} for which we distinguished
between a leading and a trailing component we did indeed find that the lower
frequency micropulses arrived first in the leading component. However, we did not
observe a sign reversal of the delay for the micropulses in the second
component.

This complication in the interpretation notwithstanding, it
is conceivable that the smallest observable angular scale of the fluctuating
radiation pattern of the beam is limited by the geometry of the opening
cone of emission and therefore likely by the opening
polar magnetic field lines, the particular radius to frequency mapping, and
the bandwidth of the receiver. Shorter pulses in these pulsars may become
observable by using a smaller bandwidth. The shortest pulses observable at 1.7~GHz would
be limited by the rise time of the optimal filter to about $0.5~\mic$. Other
pulsars may have a more favorable, weaker dependence
of the ACF width on frequency and could show less angular smearing.
A detailed study of the correlation and the pulse time of
arrivals of micropulses at several close frequencies with a similar resolution
of 62.5~ns could provide important clues about the geometry of the magnetic
field lines, the elementary pulse structure, and in turn the emission mechanism.

However, the alternate interpretation of the short micropulses being due to a
radial modulation of the radiation pattern is not necessarily disfavored. In the context of
this interpretation the values for $\gamma$ could be easily two to three orders
of magnitude lower and the typical radial length of the modulation $\sim
c\cdot\tau_\mathrm{\mu-narrow}$ (see Table~\ref{psrpar3}). The sources of such short micropulses
have brightness temperatures of $\sim 10^{34}$ K.  With the influence of
beaming being excluded, the high degree of coherence of the modulation over the
radial length would primarily account for the high brightness temperature.

Several possible emission mechanisms have been proposed to explain radio
emission of pulsars. In general, they can be classified into three groups: emission by
bunches, plasma instabilities, and maser emission~\citep{melrose1992}.
Theoretical predictions on the possible generation of ultra short timescale intensity
fluctuations in the pulsar radio emission are based on nonlinear temporal models
describing the interaction of a high energy beam of particles with plasma wave
packets in the pulsar magnetosphere. \citet{asseo1990} suggested
that a two-stream instability may result in the growth of strong plasma
turbulences which would lead to a self-modulation instability, the creation of
localized wave packets, and the generation of stable Langmuir soliton-like
solutions. The latter would result in the so-called ``Langmuir microstructures''
with timescales in the range of several microseconds.
Our shortest micropulses have the predicted typical durations and could be
interpreted in this way.

Other interpretations of microstructure are based on nonlinear models of
self-modulational instabilities of an electromagnetic
wave with a large amplitude propagating in an electron-positron
plasma~\citep{chian1983, onishchenko1990, chian1992, gangadhara1993}. The models
demonstrate that the modulation instability can evolve to a nonlinear
stationary state that results from the balance between nonlinearity and
dispersion. The possible stationary solutions of the respective nonlinear
equations are envelope solitons or isolated wave packets and periodic wave trains.
The first solution may describe
strong isolated micropulses while the second could describe the
phenomenon of quasi-periodicity.

Recently~\citet{weatherall1998} conducted a numerical simulation based on
nonlinear wave dynamics which provides detailed solutions for the temporal
behavior of pulsar radio emission. The model predicts an intrinsic pulse width in
the  range of 1--10~ns (nanostructure). The characteristic timescale is
expected to be frequency dependent, becoming longer at lower frequencies.

Our analysis of pulsar radio emission with the resolution of 62.5~ns
has not detected any nanopulses. Nanostructure may in general
be present in radio emission as ``shot noise''
as suggested by~\citet{cordes1976b}. Perhaps nanopulses might be
observed as an infrequent phenomenon like the ``giant'' pulses from the Crab
pulsar.

However, no matter how short elementary pulses could intrinsically be, they
are not only likely affected by angular smearing as suggested for our pulsars
above but could perhaps also be  significantly affected while propagating through the
magnetosphere toward the observer. \citet{lyutikov2000} have studied the
scattering and diffraction  of radio waves in the
pulsar magnetosphere and found that the  scattering
delay can be as large as $10~\mic$, easily explaining our cutoff widths in the
histograms. The dispersion delay of micropulses of about $2~\mic$
we found for \object{PSR~B0950+08} and both components of \object{PSR~B1133+16}
could also be explained in their model as a propagation effect.

The difference in the characteristics of microstructure for the two components
of \object{PSR~B1133+16} remains puzzling. If the two components define the
longitudes where the hollow cone of emission is intersected by the line of sight, then
the emission mechanism produces micropulses with characteristics as a function
of azimuth about the axis of the hollow cone. Perhaps asymmetries in the
polar cap magnetic field structure could cause azimuthally dependent plasma
bunching properties and propagation effects that would result in the differences
we observe.

\section{Conclusions}

The main results and the conclusions of our study can be summarized as follows:
\begin{enumerate}

\item The S2 VLBI system provided a conceptually clearly defined continuous
flow of data with a time resolution of 62.5~ns that could be transferred
via the S2-TCI to small digital tapes for the analysis of pulsar radio signals.

\item Broad microstructure was found for the first
time in the average CCF (or ACF) of \object{PSR~B1929+10} and confirmed for
PSRs \object{B0950+08} and \object{B1133+16}.

\item Narrow microstructure with a typical width of only about $10~\mic$
was discovered in the average CCF of PSRs \object{B0950+08},
\object{B1133+16} (component II) and \object{B1929+10}.

\item Histograms of microstructure timescales show a steep increase toward
shorter timescales followed by a sharp cutoff at about 5--10~$\mic$.
The shortest micropulse detected had a width of $2~\mic$.

\item No unresolved nanopulses or pulse structure with a sub-microsecond
timescale were found. This conclusion follows from the comparison of histograms
of amplitudes for ON-pulse and OFF-pulse windows; it is confirmed by the completely
identical shapes of the high-frequency portions of the average cross-power
spectra for ON-pulse and OFF-pulse emission; and it is supported by the 
short-term ACFs which show that the pulsar signal is indistinguishable 
from pure receiver noise at timescales shorter than $8~\mic$.

\item For \object{PSR~B1929+10} the dispersion delay for micropulses over 16~MHz
or $\sim 1\%$ of the center frequency is the same
within one standard error as that for the average pulse over a much larger
absolute and relative frequency spread. For PSRs \object{B0950+08} and \object{B1133+16}
the residual delay is about $2~\mic$ with the micropulses at the lower
frequency arriving earlier than expected.

\item The cross-power spectra for single pulses
display features with an almost continuous variety of complexity
ranging from single features that correspond to the classic quasi-periodicities to
multiple and broad overlapping features with random positions in the
spectra that correspond to essentially no quasi-periodicities at all.

\item  The power spectral density of the average cross-power spectra of single
pulses decreases steeply toward higher frequencies of the microstructure
quasi-periodicities and merges with the power spectral density of the OFF-pulse emission
at a frequency that corresponds to the shortest microstructure
quasi-periodicities.

\item The statistics of the micropulses and their quasi-periodicities differ
significantly for the two components of \object{PSR~B1133+16}.

\item It is conceivable from our data that the shortest micropulses  represent,
at least in part, the shortest angular scale of the longitudinal radiation pattern
of the sweeping beam with $\gamma \sim 20\,000$ for the radiating particles.
The particular combination of the geometry of the opening, curved magnetic
filed lines, the radius to frequency mapping, and the bandwidth over which the
radiation is observed limited the observable width of micropulses in our
observations to $\sim 5~\mic$. The shortest micropulses
of our three pulsars at 1.7~GHz are predicted to have a width of
$\sim 0.5~\mic$. If the short micropulses are instead due to a radial
modulation of the radiation pattern the typical length of the pattern is $\sim 3$ km.

\end{enumerate}

\begin{acknowledgements}
We thank the anonymous referee for very helpful suggestions for an improvement of the 
paper.
M.V. Popov thanks the Space Geodynamics Laboratory at CRESTech and York University
for providing support for his work on the data reduction while he was staying in
Toronto. N. Bartel thanks the Astro Space Center of the Lebedev Physical Institute in Moscow,
the Canadian Institute for Theoretical Astrophysics (CITA) in Toronto, and
the Observat\'orio Nacional in Rio de Janeiro for their hospitality and support
during part of his sabbatical year while this paper was being written. This investigation
was supported in part by the Russian Foundation for Fundamental Research (project's numbers
98-02-16917 and 01-02-16871), by INTAS (project number 96-0154), and by Canada's
NSERC. The DSN 70-m telescope is operated by JPL/Caltech under contract with the
National Aeronautics and Space Administration.\\

\end{acknowledgements}

\appendix
\section{Correction of the average CCF for the effects of receiver noise}
\label{appA}

The effects of receiver noise on the average ACF of a pulsar signal were
investigated by~\citet{rickett1975}. The ACF, $r(\tau)$,
of the pulse intensity in terms of observable quantities is given as
$$
   \langle r(\tau)\rangle =
   \langle r_\mathrm{on}(\tau)\rangle - \langle r_\mathrm{off}(\tau)\rangle
   \left(1+2\frac{\langle I_\mathrm{on} \rangle}{\langle I_\mathrm{off} \rangle}\right)~,
$$
where $r_\mathrm{on}(\tau)$ and $r_\mathrm{off}(\tau)$  are the ACFs for the ON-pulse and
OFF-pulse windows each averaged over all pulses. Further, $\langle I_\mathrm{on}\rangle$ 
is the mean intensity in the respective window 
again averaged over all pulses, but with  $\langle I_\mathrm{off}\rangle$ subtracted.  
The normalized ACF, $R(\tau)$, is then given as
$$
   \langle R(\tau)\rangle =
   \frac{\langle r(\tau)\rangle}{\langle r(0)\rangle} =
   \frac{\langle r_\mathrm{on}(\tau)\rangle}
        {\langle r_\mathrm{on}(0)\rangle - \langle r_\mathrm{off}(0)\rangle \left(1+2\frac{\langle I_\mathrm{on} \rangle}{\langle I_\mathrm{off} \rangle}\right)}~,
$$
$\langle r_\mathrm{off}(\tau)\rangle$ being essential zero for $\tau > 1/\mathrm{B}$, where $\mathrm{B}$
is the receiver bandwidth.

This equation can be rewritten in terms of
$\langle R_\mathrm{on}(\tau)\rangle = \frac{\langle r_\mathrm{on}(\tau)\rangle}{\langle r_\mathrm{on}(0)\rangle}$ as:
$$
   \langle R(\tau)\rangle =\frac{\langle R_\mathrm{on}(\tau)\rangle}{\epsilon}~,
$$
where $\epsilon$ is a correction factor:
$$
   \epsilon = 1 - \frac{\sigma_\mathrm{off}^2}{\sigma_\mathrm{on}^2}\left(1+2\frac{\langle I_\mathrm{on} \rangle}{\langle I_\mathrm{off} \rangle}\right)
$$
with $\sigma_\mathrm{on}=\sqrt{\langle r_\mathrm{on}(0)\rangle}$ and $\sigma_\mathrm{off}=\sqrt{\langle r_\mathrm{on}(0)\rangle}$.

In this paper we used the CCF between adjacent 16-MHz frequency channels. It can be seen that for 
the CCF the correction
factor $\epsilon$ should be modified to:
$$
   \epsilon = 1 - \frac{\sigma_{\mathrm{off}_1}}{\sigma_{\mathrm{on}_1}}\frac{\sigma_{\mathrm{off}_2}}{\sigma_{\mathrm{on}_2}}
   \left(1+\frac{\langle I_{\mathrm{on}_1} \rangle}{\langle I_{\mathrm{off}_1} \rangle} + \frac{\langle I_{\mathrm{on}_2} \rangle}{\langle I_{\mathrm{off}_2} \rangle}\right)~.
$$

In fact, in our observations, the values of $\sigma_\mathrm{on}$, $\sigma_\mathrm{off}$,
$\langle I_\mathrm{on}\rangle$, and  $\langle I_\mathrm{off} \rangle$ in one channel never differed
from the equivalent ones by more than a few percent in the other channel
because of the proximity of the two frequency channels and the AGC operation.

\bibliographystyle{apj}
\bibliography{biblio}

\end{document}